%% file: s0_mn.tex
\documentclass[usenatbib]{mn2e}

\usepackage{graphicx}

\def\simless{\buildrel < \over \sim}

\def\less{< }

\def\arcsc{\char'175 }

\def\etal{et~al.\ }

\def\kms{~km~s$^{-1}$}
\def\mpc{~Mpc$^{-3}$}

\title[Stellar Population Trends in S0 Galaxies]{Stellar Population Trends in S0 Galaxies}

\author[L.C. Chamberlain] {L. C. Prochaska Chamberlain$^{1}$\thanks{Email:lchamberlain@sps.edu}\thanks{Present address: St. Paul's School, Concord, New Hampshire 03301} St\'{e}phane Courteau,$^2$ Michael McDonald,$^3$ James A. Rose$^1$ \\$^1$Department of Physics and Astronomy, CB 3255, University of North Carolina, Chapel Hill, NC 27599 \\$^2$Department of Physics, Engineering Physics and Astronomy, Queens University, Kingston, ON, Canada \\$^3$Department of Astronomy, University of Maryland, College Park, MD}

\begin{document}
 
 \maketitle

\begin{abstract}
We present stellar population age and metallicity trends for a sample of 59  
S0 galaxies based on optical SDSS and NIR J \& H photometry. When
combined with optical { \it g}  and  {\it r} passband imaging data from the SDSS
archive and stellar population models, we obtain radial age and metallicity 
trends out to at least 5 effective radii
for most of the galaxies in our sample. The 
sample covers a range in stellar mass and light
concentration. 
We find an average central light-weighted age of $\sim$ 4 Gyr and central 
metallicity [Z/H] $\sim$ 0.2 dex. Almost all galaxies show a negative metallicity 
gradient from the center out, with an average value of
$\Delta$[Z/H]$/$$\Delta$log(r/$R_e$) $=$ -0.6. An age increase, decrease, and minimal change with radius is observed for  58\%, 19\%, and 23\%, respectively, for a mean age gradient of $\Delta$age$/$$\Delta$log(r/R$_e$) $=$ 
2.3 Gyr dex$^{-1}$. For 14 out of 59 galaxies, the light-weighted age of the 
outer region is greater than 10 Gyr.   We find that galaxies with both lower 
mass and lower concentration have younger light-weighted ages and lower light-weighted metallicities. This mass-metallicity relation extends into the outer regions of our S0 galaxies. Our results are consistent with the formation of S0 galaxies through the transformation of spiral galaxy disks. Determining the structural component that makes up the outer region of galaxies with old outksirts is a necessary step to understand the formation history of S0 galaxies.

\end{abstract}

\begin{keywords}
galaxies: elliptical and lenticular, cD; galaxies: evolution; galaxies: formation; galaxies: photometry
\end{keywords}

\section{Introduction}

The unique morphological status of S0 galaxies between gas-poor and gas-rich 
galaxies makes their origin a key ingredient of
galaxy formation and evolution models. Numerous hypotheses
suggest that at least a class of S0 galaxies are by-products of secondary 
events in galaxy formation that transform spiral galaxies into S0 galaxies, 
rather than products of initial conditions  
\citep{Ick85, BV90, Moo96, Qui00, Bos08}.  These
theories involve the removal of gas and subsequent halting of star formation
and can include the following formation mechanisms: minor interactions with
other galaxies \citep{Ick85, Bar02}, harassment \citep{Moo96}, ram pressure
and/or viscous stripping \citep{Qui00, Kro08}, strangulation \citep{KM08}, and
cluster tidal effects \citep{BV90}.
Observational evidence for a higher
fraction of S0-to-spiral galaxies in local clusters as compared to 
higher redshift clusters certainly supports the idea that S0 galaxies are
primarily created through evolutionary processes \citep{BO78, DG83}. 
However, obstacles to 
theories of a simple transformation from a spiral to a lenticular galaxy exist.
For example,  the bulge-to-disk ratios of S0 galaxies appear to be on 
average larger than those of spirals in all density regimes \citep{Dre80},
thus creating problems for a straightforward disk-quenching scenario, although not
for other scenarios such as merging. 
The structural dichotomy between massive and low mass S0s
suggests different formation scenarios for the two mass groups
\citep{Van94, Bed08}.  In this 
paper we study the stellar populations in S0 galaxies to better 
constrain their possible formation processes.

Observations of radial stellar population (hereafter, SP) gradients in S0 galaxies
may help to distinguish between possible scenarios for S0 galaxy formation, since
the dynamical events that lead to the production of S0s may produce or erase 
imprints on the galaxy's star formation history.
For example, hierarchical models of galaxy formation predict an inside-out
accretion of disk gas from the hot gaseous halo, resulting in ages that
decrease outward in the galaxy (Fall \& Efstathiou
1980, but see Renzini 1995).
On the other hand, formation mechanisms, such as those outlined above for the transformation of a spiral galaxy into an S0, could alter this trend of age with radius through either recent star formation \citep{BV90,Moo96,Qui00,Kro08} or gas removal \citep{Qui00,Kro08}. The latter process can alter age trends with radius because gas is preferentially removed from the outer disk, while for the former, this occurs through gas inflow.

SP trends in the inner regions of S0 galaxies have been widely studied. Spectroscopic studies suggest that the nuclear regions of S0 galaxies are on average younger and more metal rich than bulges  \citep{Fis96, Sil06, Ser08}; however, \citet{Ser08} found this to be true only for HI-poor S0 galaxies. Although the photometric studies of \citet{Pel99} found redder colors in the nuclear regions of S0s compared to further out, they attributed this to dust 
effects, as opposed to age differences.  Most models for the transformation of spiral into S0 
galaxies predict a resulting central gas concentration that leads to a young 
SP in the center of the S0 remnant \citep{Ick85,Moo96, Kro08}, 
consistent with spectroscopic studies (as cited above). 
Derivations of the mean central ages yield an uncertain picture. 
\citet{Pel99} examined the bulges of S0 galaxies using Hubble Space 
Telescope images and found 
an average bulge light-weighted age of 9 Gyr, while \citet{Sil06} found ages between 
 4.8 and 8.3 Gyr. Both studies find high central metallicities. 

Even less is known about the age trends at larger radii in S0 galaxies. 
Various authors have proposed that the disks of S0 galaxies are 
younger than the bulge \citep{Cal83,BG90,Nor06}. 
Others suggest comparable ages for the bulge and disk
\citep{Pel96, Meh03} while some report older ages in the disk compared to 
the bulge  \citep{Fis96,Tik03,Mac04}. The large variation in observed age trends may be due to differences in radial coverage, bandpasses studied, and the adopted SP  
models, if any, by the authors.  Alternatively, it is possible that the disagreement between studies
may be due to the heterogeneous nature of S0 galaxies. 
Large samples with accurate photometry extending 
beyond a few effective radii are required to firm up SP and evolutionary trends in  
S0 galaxies.

In addition to establishing radial SP gradients in S0 galaxies,
one must assess whether the SP gradients and mean values are linked to global
properties such as the galaxy's stellar mass, central light concentration, and environment. 
S0 galaxies are proportionately more populous in the dense environments of rich clusters than in low-density environments \citep{Dre80}, suggesting that environment plays a role in S0 formation and evolution. The mass of a galaxy is closely tied to its formation history. Hierarchical assembly predicts that larger galaxy dark matter halos are
formed through mergers later in time than smaller halos \citep{NFW95}.
However, semi-analytic models can
produce ``anti-hierarchical'' star formation histories in a $ \Lambda$ Cold Dark Matter ($ \Lambda CDM $) 
universe despite the hierarchical assembly of these galaxies \citep{del06}. 
As well, \citet{Nei06} showed analytically that anti-hierarchical evolution can be a 
 natural outcome of bottom-up clustering. 
Such anti-hierarchical evolution is more consistent with actual observational
trends of stellar populations with galaxy mass for all Hubble types, including 
S0s \citep[][hereafter, M04]{Mac04}.
For example, an increasing fraction of S0 galaxies with recent star 
formation has been observed for decreasing luminosities (an indicator of 
stellar mass) \citep{BP94,Pog01}. Additionally, a correlation of older stellar 
populations in S0 galaxies with higher velocity dispersion (an indicator of 
dynamic mass) has been observed \citep{Meh03,Sil06,Bed08}. These studies 
support different star formation histories for high and low mass S0 galaxies, 
but have largely focused on the inner stellar populations of 
galaxies. 

The degree to which light is concentrated towards the center of the galaxy is also 
linked to a galaxy's formation history. Secondary galaxy evolution 
processes may increase the mass of the central region
either through secular build-up or accretion of satellites 
\citep{Kor04}. On the other hand, the relative mass of the central component, 
or bulge, may influence the outcome of galaxy formation processes, such as 
those transforming spiral galaxies. For example, harassment preferentially 
selects bulge-dominated galaxies to be stripped of their spiral structure and 
transformed into S0s, since disk-dominated galaxies will instead be shredded down to a 
dwarf system \citep{Moo96}. Either way, the light concentration of the 
galaxy is closely tied to galaxy evolution processes and should thus correlate 
with SP parameters. For instance,  \citet{M09a} found a lack of
low surface brightness galaxies with high concentration as well as a lack of high
surface brightness galaxies with low concentration.
Therefore, they suggest that galaxies with high concentration can only 
form through major mergers.  Other studies relating the size of the bulge 
and the bulge-to-disk ratio (which is related to concentration) to the evolution
of S0 galaxies find that galaxies with larger 
bulges are redder and have smaller color gradients \citep{BP94}. However, 
at intermediate redshift (0.73 $<$ z $<$ 1.04), \citet{Koo05} found no difference in bulge colors between
disk and bulge-dominated galaxies, suggesting that B/D
ratio is poorly correlated with stellar populations. 

Despite recent progress, studies of 
radial SP trends in S0 galaxies as a function of the global properties of mass and concentration are still open-ended and have seldom explored
the regions beyond $1-2~R_e$. Most efforts have indeed focused on age
and chemical composition within one effective radius.
In this paper, we use deep optical/near-infrared (NIR) color gradients to explore age and metallicity
gradients in S0 galaxies out to typically $5~R_e$, and with a
large enough sample to cover a substantial range in galaxy mass and light
concentration.
The focus of this paper is largely on empirical radial trends in 
stellar populations; separate bulge and disk trends that result from model
decompositions of the galaxy image will be presented in a forthcoming publication (Chamberlain et al. 2011, in prep).
Our sample of 59 S0 galaxies is presented in \S\ref{sect:sample}, while our observations and data 
reduction techniques are discussed in \S\ref{sect:obs} and  \S\ref{sect:data}, 
respectively. Measurements of global galaxy properties are presented in 
\S\ref{sect:param} and the SP models are described in \S\ref{sect:stelpop}. 
Our results for the age and chemical composition trends in S0 galaxies are
presented in \S\ref{sect:results}.  The implications 
of our results are discussed in \S\ref{sect:discussion}, and a summary 
is presented in \S\ref{sect:conclusion}. A distance of 16.5 Mpc is used for all Virgo cluster 
galaxies (Mei et al. 2007).

\section{Sample}\label{sect:sample}

We selected from the UGC catalog \citep{Nil73} all galaxies with 
S0, S0a and S0B 
morphologies and blue Galactic extinction {$ \leq 0.5$} yielding a 
catalog of 1088 objects. This
sample was further restricted to areas of the sky covered by the Sloan Digital
Sky Survey \citep[][hereafter, SDSS]{Yor00} bringing the sample to 542 
galaxies. Finite telescope time limited us to 
NIR data for only $\sim$ 15\% of the SDSS sub-sample; the sample was selected to cover a representative range of mass and light concentrations. Eliminating edge-on galaxies and galaxies with significant spiral structure or overlapping foreground stars further restricted our sample to 59 objects as described below. While this UGC-derived sample includes some range in environment (see \S\ref{sect:param}), the higher density environments are missed. Thus, for greater coverage, we include observations of eight low-mass Virgo cluster galaxies from the sample of \citet{M09b} (hereafter M10); these were observed with the same equipment as our 
sample (see \S\ref{sect:obs}). However, these eight galaxies all have low mass and low concentration; see \S\ref{sect:param} for more details. In the electronic version, we show SDSS color images (using $\emph{g}$, $\emph{r}$, and $\emph{i}$ bands) for our full sample. These images were obtained from the SDSS website:  http://cas.sdss.org/dr7/en/tools/chart/navi.asp.

Our sample excludes galaxies with spiral structure and highly disturbed structure, 
as gauged by visual inspection of the SDSS images. Galaxies with tidal tails or faint 
spiral structure were retained, but classified as 
 ``transition'' galaxies due to the emergence or fading of their spiral structure. Examples are shown in the color images available in the electronic version, and are noted by a ``T'' above the image.  Twelve of our 59 sample 
galaxies are labeled as ``transition''. 
We will see in \S\ref{sect:agetrend}  that the putative structural transition 
correlates with a transition in SP properties. We note that some
S0 galaxies left in our ``featureless disk'' sample are likely to reveal 
spiral structure when examined with higher resolution images and our sample may also contain 
elliptical galaxies misclassified as S0. 

Highly inclined galaxies with {$ i \geq 75$} $\deg$, and galaxies that fall 
on the edge of the SDSS field of view, were also pruned from the sample
in order to measure well-defined surface brightness (SB) profiles. 

Our final sample is neither statistically complete 
nor randomly selected, but tailored to cover a broad range in global properties such as 
mass, concentration, and local environmental density (see \S\ref{sect:param}). 
The galaxy sample is presented in Table 1. 

\section{Observations}\label{sect:obs}

Our SP analysis relies on optical and NIR photometry of 59 S0 galaxies. 
The NIR observations (J- \&/or H-band) were collected with the ULBCam at the University of 
Hawaii's 2.2-m telescope and the archival optical imaging (\emph{g} \& \emph{r}) is 
from SDSS.

\subsection{Optical Observations}

Optical \emph{g} \& \emph{r} images with a scale of 0.396 arcsec/pixel 
were extracted from the SDSS/DR5 archives. The \emph{u} and \emph{z} bands 
were left out considering their low signal-to-noise. We did not use the \emph{i}-band 
since it does not improve the discriminating power for age and metallicity in SP 
models, when combined with \emph{g}, \emph{r}, and J or H bands. 

\subsection{Near-IR Observations}\label{sect:Near-IR}

Near-IR images for the 59 S0 galaxies in our sample were obtained with the ULBCam at the 
University of Hawaii's 2.2-meter telescope on Mauna Kea in April 2005, 
2006 and 2007 and March 2008.  
A total of 29 galaxies in our sample were observed only in the H band, 19 were 
observed only in the J band, and 11 galaxies were observed in both the 
J and H bands. 
The ULBCam image scale is 0.25 arcsecond per pixel. A 
maximum single exposure time of 40 seconds was used to 
maximize the sky flux whilst keeping within the detector's linear regime. 
The total integration time depends on SB and ranges from 8 to 24 minutes. A standard dithering script minimized the resampling of
bad pixels. Only the cleanest of the four 2048x2048 arrays was used,
resulting in an 8.5' by 8.5' field of view, which was large enough to 
properly assess the sky background levels.  
Our  NIR flux calibration makes use of reference stars in the 
target galaxy field from the Two Micron All-Sky 
Survey \citep[][hereafter 2MASS]{Skr06}. Thus, no standard star observations for 
photometric calibration were necessary (see \S\ref{sect:data2}). A more detailed description of ULBCam
 data taking procedures is provided in M10. A log of observations is presented in Table 2.  

In the remainder of the paper, we refer to the galaxies observed in the J-band as the 
``J-band sample'' and those observed in the H-band as the ``H-band sample''. 
Eleven overlapping galaxies are included in both samples.

\section{Data Reduction}\label{sect:data}

\subsection{Basic Reductions}\label{sect:data2}

SDSS images are already processed for 
flat-fielding, bias subtraction, and cosmic ray rejection. The photometric 
calibration of the light profiles (\S\ref{sect:sbandc}) used the photometric 
zero-points provided in the SDSS/DR5 
library.

Basic reductions of the UH NIR data, which include flat-fielding, stacking,
bad pixel rejection, geometric distortion corrections, and flux calibration, were
applied to the ULBCam data using the XVISTA software 
package\footnote{http://astronomy.nmsu.edu/holtz/xvista/index.html}. The data reduction 
procedures follow the prescription 
of M10. The NIR flux calibration of light profiles used infrared stellar photometry from 
2MASS for stars in the ULBCam target galaxy 
fields. A mean offset between 
our brightnesses and the 2MASS J or H-band brightnesses was calculated and 
used for the zero-point calibration. This
method of flux calibration enabled us to calibrate the photometry of our science images 
at each pixel, independent of airmass variations and transparency conditions. 
Additional information on flux calibration methods 
and background stability can be found in M10. The photometric zero-point and the standard deviation about the mean in the calibrating stars 
are noted in Table 2. If fewer than three 2MASS stars were found in a target field (indicated by a zero in the last column of Table 2), 
the calibration used is the average of the previous and next exposures. The quoted standard deviation for these observations is a standard deviation of all calibration values, that is $\pm$ 0.056 mag. 

SBs have been corrected for Galactic foreground 
extinction in each bandpass using the reddening values, $A_{\lambda}$, of \citet{Sch98} 
and assuming an $R_{V} = 3.1$ extinction curve.

\subsection{PSF Matching}\label{sect:psf}

Wavelength-dependent image blurring by the atmosphere affects 
inner galaxy colors. Thus, we
measured the 2D seeing point spread function (PSF) to correct galaxy color profiles. 
The PSF full width at half maximum
(FWHM) is measured from the final stacked 
image in each bandpass. The stars that
are identified for PSF measurements in the \emph{g} \& \emph{r} bands are 
individually sky-subtracted and fit with a
2-dimensional Gaussian function from which FWHMs are 
derived. On average, 50 stars are used for PSF measurements in each 
stacked SDSS image. The
final PSF per image is the median of all individual star PSFs.  
For NIR images, the final PSF per image is the average of 5 
individual PSFs. Images from each bandpass are degraded to the PSF of the
worst seeing bandpass (typically the \emph{g} image) by convolving with a 2D
gaussian of appropriate FWHM. 
For additional precaution, we apply an inner radial cutoff 
of roughly 2 seeing disks (3 arcseconds) to the color profiles. The central region 
in all other analyses of colors is averaged within at-least the inner 2 seeing disks. 
PSF measurements have also enabled the identification of foreground stars that 
are then removed from the galaxy light.

\subsection{Sky Measurement}\label{sect:sky}

Careful sky subtraction is crucial for the accurate analysis of deep 
SB profiles. Sky subtraction follows slightly different
approaches for SDSS and NIR images.

We have used the sky 
value provided with each SDSS 
image\footnote{http://www.sdss.org/dr5/algorithms/sky.html}. 
For the images where the 
sky value was not available ($\sim$ half our sample), we 
used the lowest of the sky measurements measured by us from either the 
entire SDSS image or 
from the four image corners. 
The latter scheme gives sky estimates within 0.5\% of the SDSS values (M10). 

For the NIR data, the sky is measured in 
four rectangular boxes along the perimeter of the field of view 
and then averaged together to give the mean sky value for that image. 
Each of the four sky boxes has a typical size of $\sim$ 75 x 500 arcseconds.  
The typical deviation of sky levels amongst the four boxes for the NIR data 
are .004\% of the sky value.

\subsection{Surface Brightness Profile Extraction and Error Estimates}\label{sect:profs}

SB profiles were extracted by fitting 
elliptical isophotes to the \emph{r}-band galaxy images. The XVISTA 
command, PROFILE, enables this operation based on a generalized non-linear 
least-squares fitting routine. For these fits, ellipticity and position angle 
are allowed to vary but the galaxy center is fixed. The isophotal solutions based on the \emph{r}-band images were then applied uniformly to the images 
in the \emph{g}, J, and H band images. This ensures that color gradients are 
computed from the same matching isophotes. Further details about isophotal fitting and profile 
extraction, including details on profile depth and signal-to-noise ratios, are given in \citet{Cou96} and M10.  

The effect of a systematic sky 
error on galaxy photometry naturally increases with galactocentric radius. 
To estimate this effect, we have recalculated the SB profiles and
color profiles using sky values
adjusted to $sky = sky_{orig}$ $\pm 1\sigma_{sys}$, where $\sigma_{sys}$ is 
the systematic sky uncertainty. This is calculated as the standard error in the mean 
of the sky values (as described in \S\ref{sect:sky}) from four boxes along 
the perimeter of the image.
We determine the sky effect at each radius along the SB profile as half the 
difference in the SB calculated with the 
sky value set at $sky^{+} = sky_{orig}$ $+$ $\sigma_{sys}$ and $sky^{-} = sky_{orig}$ - $\sigma_{sys}$. 
Our final computation of the photometric error at each radius is the sum in quadrature of the 
statistical error from profile fitting (this naturally includes statistical errors in the sky background) and 
the systematic sky error estimate.

\subsection{Outer Radial Cutoff}\label{sect:cutoff}

\begin{figure}
\includegraphics[width=90mm]{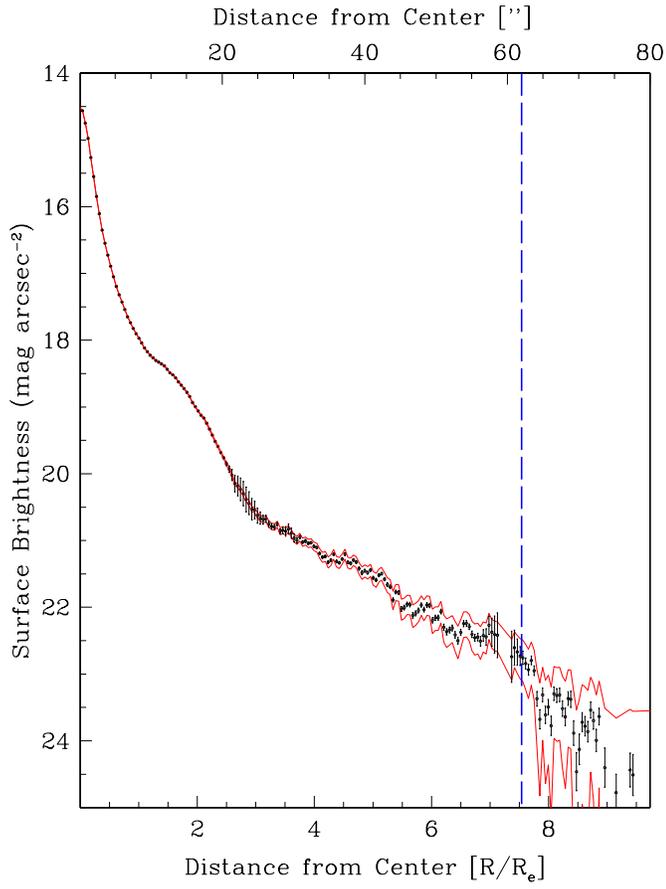}
\caption{H-band radial SB profile for UGC4737. The SB profile is shown as black dots with  $\pm \sigma$ sky error envelopes as red lines. The sky error envelopes are calculated using sky values adjusted to 
$sky = sky_{orig}$ $\pm 1\sigma_{sys}$. Black error bars are statistical SB errors from profile fitting. The upper axis indicates the radial extent from the center in arcseconds  while the lower axis is scaled by the \emph{r}-band half-light radius. The outer cutoff is shown as a vertical blue dashed line.  }
\label{fig:skyerrenv}
\end{figure}

The low NIR galaxy SB, relative to the
bright NIR sky background, make an accurate determination of the sky
background crucial for extracting reliable surface photometry.
This is demonstrated in Fig.~\ref{fig:skyerrenv} for a 
sample galaxy (UGC4737).  We
define the outer radial cutoff as the point where the sky error envelopes deviate from each other by more than 0.4 mag arcsecond$^{-2}$, which is usually where the statistical error per point is comparable to the typical point-to-point variations. Our NIR profiles are intrinsically shallower 
than SDSS profiles and the common outer truncation radius was thus determined using 
J and H profiles for all but UGC 10158 where the optical bandpasses are more sensitive 
to sky errors. In the few cases where the statistical error from profile fitting (for any of the optical or NIR bands) consistently exceeds 0.1 mag arcsecond$^{-2}$ at 
large radii, the radius  at that SB threshold point would be used as the outer cutoff instead.
An example of the outer cutoff, based on an H-band profile, is shown in 
Fig.~\ref{fig:skyerrenv} as the dashed vertical line. The radial cutoff used for the \emph{g} and 
\emph{r} bands matches the NIR band that is being used for analysis. The outer radial 
cutoffs for each galaxy in the J and H-bands are given in 
Table 3. 

\subsection{Surface Brightness and Color Profiles}\label{sect:sbandc}

\begin{figure}
\includegraphics[width=90mm]{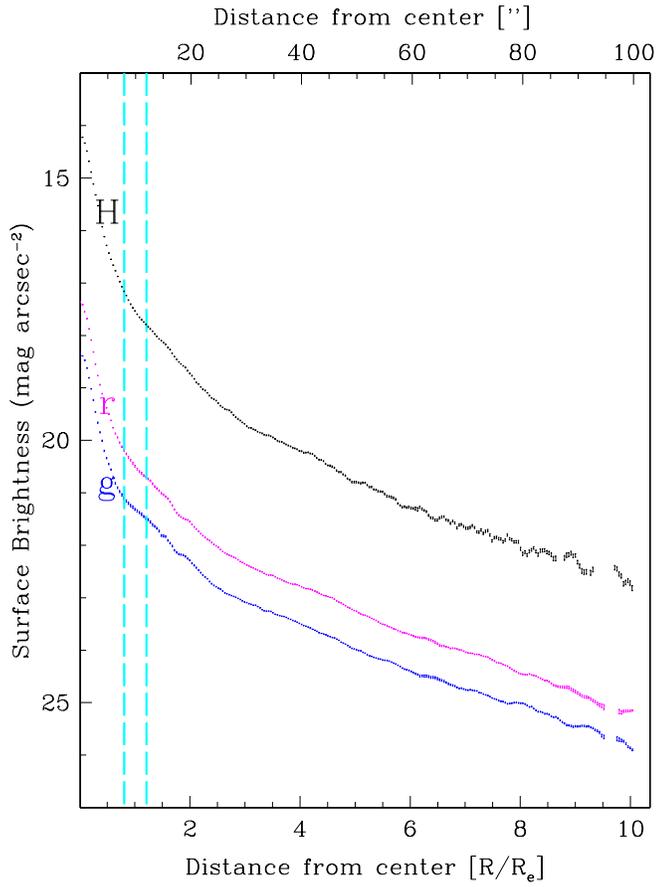}
\caption{Radial SB profiles in H-band (black), \emph{r}-band (magenta), and \emph{g}-band (blue) for UGC 4869. The vertical dashed cyan lines indicate the separation between the inner and outer regions. 1 $\sigma$ statistical SB errors are shown at each radial point.}
\label{fig:sbprofex}
\end{figure}

\begin{figure}
\includegraphics[width=90mm]{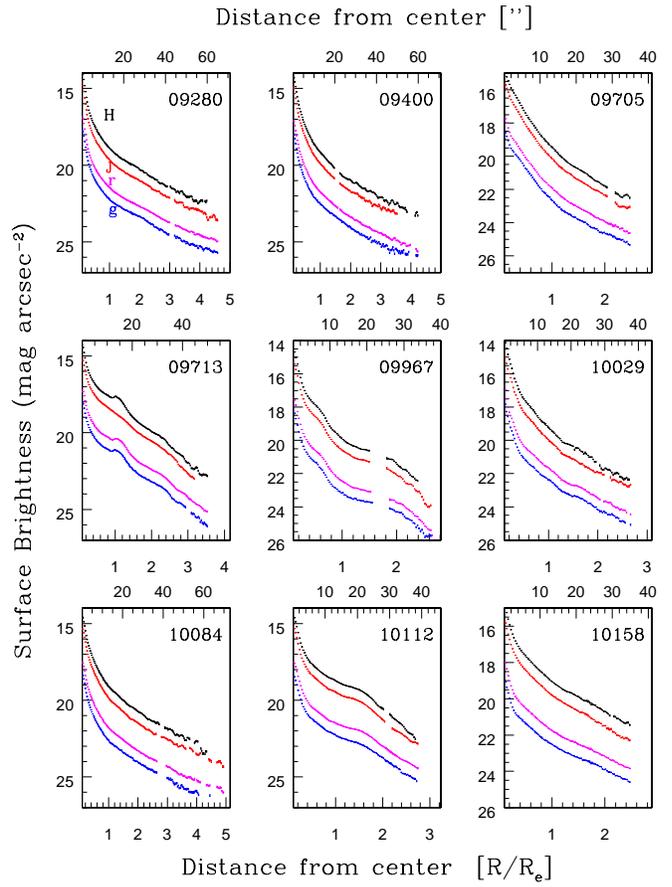}
\caption{Radial SB profiles in \emph{g}-band (blue), \emph{r}-band (magenta), J-band (red), and H-band (black) are shown. UGC or VCC numbers of the galaxies are given in the upper right corner of each panel. [$\emph{See the electronic edition of the Journal for the remainder of our sample.}$]}
\label{fig:sbprofs}
\end{figure}

\begin{figure}
\includegraphics[width=90mm]{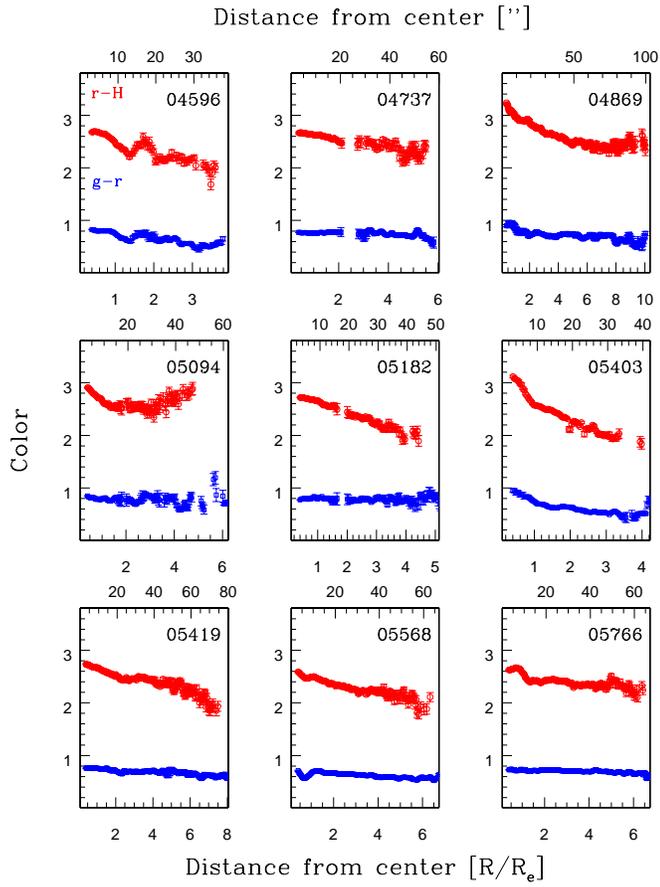}
\caption{Color profiles in \emph{r}-H (red) and \emph{g}-\emph{r} (blue) are shown. The lower axis indicates the radius in terms of the \emph{r}-band half-light radius ($R_e$) while the upper axis is in units of arcseconds. The $\pm$1 $\sigma$ error bars are computed in quadrature from each of the combined SB profiles.  UGC or VCC galaxy numbers are shown in the upper right corner of each panel. [$\emph{See the electronic edition of the Journal for the remainder of our sample.}$]}
\label{fig:colprofs}
\end{figure}

Optical (\emph{g} \& \emph{r}) and NIR (J and/or H) SB  
profiles are extracted for all of our galaxies following the prescription 
outlined in \S\ref{sect:profs}. Fig.~\ref{fig:sbprofex} shows an example surface
brightness profile for UGC 4869.  The entire collection of SB profiles 
is shown in 
Fig.~\ref{fig:sbprofs}. The \emph{g}, \emph{r}, J and H bandpasses are represented 
by blue, magenta, red and black data points.
The J and H profiles are displayed for the 11 galaxies observed in both
passbands.
Color profiles for \emph{g-r} and \emph{r}-H versus radius are shown in Fig.~\ref{fig:colprofs}. In figures where both J and H profiles are displayed, the cutoff for 
the \emph{g} and \emph{r} passbands is determined by the 
larger of the J and H-band cutoffs.  The SB and color 
error bars at each 
radius reflect the $\pm$ 1$\sigma$ statistical errors only. These errors tend to be small,
generally smaller than the point size, except at the largest radii. Interior to the outer radial
cutoff, SBs and colors are not plotted for radii where the statistical error exceeds 0.1 mag arcsec$^{-2}$, which is equivalent  to setting very low weight for these pixels. Gaps in SB and color profiles, such as that seen for UGC04737 around 25", are
due to overlapping foreground stars or galaxies. Since these foreground objects have been masked during ellipse fitting, which is identical to setting very low
weight for these pixels; 
they do not affect the profile shape. 

We compute effective radii and total magnitudes
in the \emph{g}, \emph{r}, J, and H bands.  To determine the total magnitude, we 
have extrapolated the SB profile outward by fitting the 
outer profile with an exponential 
function.  See \citet{Cou96} for more details. The 
effective radius, R$_{e}$, is the radius that contains 50\% of the 
total extrapolated light.  
For the 5 galaxies with a plateau in their SB profiles \citep[Type II;][]{Fre70}, no extrapolation is performed.
Total magnitudes, mean colors and R$_{e}$ for each galaxy in 
our sample are given in Table 4.

The lower axis of the SB and color 
profiles in Figs.~\ref{fig:sbprofs} - \ref{fig:colprofs} shows the radius 
in terms of the \emph{r}-band R$_{e}$ of the
galaxy. Most profiles extend past 5 R$_{e}$. It is 
also apparent that our sample galaxies cover a range in profile 
shapes, suggesting a variety of contributions from galactic components. Dips and
plateaus in the SB profiles, such as those seen for UGC 4596, are due to structure 
in the galaxy, usually rings, bars, or spiral structure, and/or extinction
by dust.

\subsection{Radial Binning}\label{sect:bin}

To increase the signal per color bin,
we average colors in six radial bins, scaled by the \emph{r}-band half light radius. 
The binning scheme is 
noted in Table 5.  If the statistical SB error at any radial point exceeds 0.1 mag arcsec$^{-2}$, 
that radial point or region is excluded from the analysis. We also ensure that 
each binning region contains at least 5 radial points, otherwise that region is
excluded from further analysis. 

Each galaxy is also subdivided into an inner and outer radial region.
The separation between the two 
regions is chosen at the most prominent inflection point in the SB profile.  We 
find a well-defined change in the slope of the SB profile 
for $\sim$ 50\% of 
the sample, with the majority having a break between 0.8 and 
1.2 R$_{e}$. We thus label inner regions as those within 0.8 R$_{e}$ and outer regions 
as those beyond 1.2 R$_{e}$ for all galaxies in our sample. 
Fig.~\ref{fig:sbprofex} shows 
an example of the inner and outer regions with dashed lines at 0.8 and 
1.2 R$_{e}$. 
Note that our distinction of inner 
and outer regions is independent of model fitting of galactic 
components (i.e. disk and bulge). Analysis of stellar populations of discrete 
galactic components will be discussed elsewhere (Chamberlain \etal 2010, in preparation). 
Here we focus on a model-independent analysis of the SB
profiles. For clarity, when discussing the central-most colors, we refer to 'central' as the binning region with r $<$ 0.5 R$_{e}$ and 'inner' for r $<$ 0.8 R$_{e}$.

\section{Determination and Range of Galaxy Properties}\label{sect:param}

Our analysis of SP gradients relies on correlations with global galaxy properties, such 
as stellar mass, light concentration and environment.  Prescriptions for determining
these properties
are given below and the values for each galaxy in our
sample are listed in Table 6. 

\begin{figure}
\includegraphics[width=90mm]{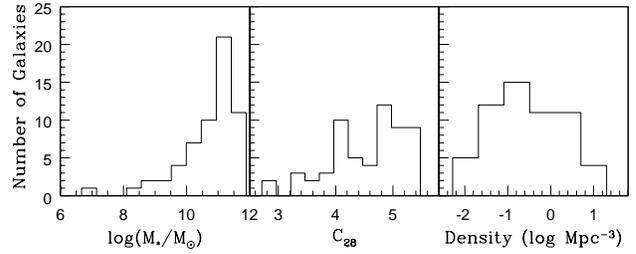}
\caption{Left: Histogram of total stellar masses for 59 galaxies.  Middle: Histogram of the galaxy light concentration, $C_{28}$, for 59 galaxies. Right:  Histogram of environmental number densities (in log \mpc) for 59 galaxies.  }
\label{fig:prophist}
\end{figure}

Total stellar masses for our galaxies were calculated from total \emph{g} and \emph{r}
magnitudes (described in \S\ref{sect:sbandc}) and using the mass-to-light ratio 
prescriptions of \citet{Bel03}, unless otherwise noted\footnote{We have also calculated masses using \citet{Por04} and find no difference in our results}. 
Distance estimates, which are required to calculate the mass and physical size of the galaxies, were corrected for Virgo flow and the Great Attractor as provided by the 
NED\footnote{The NASA/IPAC Extragalactic Database (NED) is operated by the Jet 
Propulsion Laboratory, California Institute of Technology, under contract with 
the National Aeronautics and Space Administration} for all galaxies other than 
Virgo cluster galaxies. The median stellar mass of our sample is 
1 x 10$^{11}$ $M_{\sun}$. Fig.~\ref{fig:prophist} shows a slight bias towards high mass S0s.

Model-dependent bulge-to-disk ratio estimates may carry large systematic 
errors due to the subjectivity of profile fitting functions \citep{Mac03,M09c}. 
Alternatively, the galaxy light concentration parameter gives a non-parametric indication of
the bulge-to-disk ratio \citep{Ken85}. It is computed as:
\begin{displaymath}
C_{28} \equiv 5  \log (r_{80}/r_{20})
\end{displaymath}
where $r_{80}$ and $r_{20}$ are the radii that enclose 80\% and 20\% of the 
total light, respectively. The total magnitudes and subsequent 
80\% and 20\% radii have been calculated as described in \S\ref{sect:sbandc}. 
Concentration values for our sample galaxies range from $C_{28}$ = 2.8 to 
5.5  with a median value of 4.7 as shown in Fig.~\ref{fig:prophist}. For reference, 
a pure exponential disk has $C_{28}$  $\sim$ 2.8. 

Our definition of local environment uses a 
three-dimensional number density based on the mean distance of the six
nearest neighbors.  To construct the density field, we use the Updated Zwicky Catalog \citep[][hereafter UZC]{Fal99} which is 95\% complete to a limiting magnitude of $m_{Z_{w}}$ $=$ 15.5 mag. A three-dimensional position in a Cartesian coordinate system of each galaxy in our sample is then assigned, based on its sky position and recessional velocity (using a value for
the Hubble constant of H$_{o}$ $=$ 75 \kms Mpc$^{-1}$).
The local number density of our S0 galaxies is calculated by taking the number of objects contained within a sphere of radius equal to the mean distance of the six nearest neighbors in the density field and then dividing by the physical volume of the sphere.  We note several sources of uncertainty in our environmental density estimate. 
First, the magnitude limited catalog samples the galaxy luminosity function to different levels at different heliocentric distances.  To correct for this effect, our densities are multiplied by a luminosity function (LF) correction factor, which is the ratio of the integrated observable LF (i.e., the number of galaxies per \mpc brighter than the limiting magnitude of the sample) at 3000 \kms \ to the integrated LF at the galaxy's redshift.  We use a Schechter LF \citep{Sch76} derived from the UZC, with $\alpha$$=$-1.0 and $M^*_g$$=$-18.8 \citep{Mar94}.
However, for objects with 
cz $> $ 9000  \kms , correction factors become unreliable 
(greater than factors of 3).
Fortunately, only 8 out of our 59 galaxies fall beyond this redshift range. 
We also consider that galaxies with cz $< $ 3000  \kms \ (17 galaxies in our sample)
have large peculiar velocities that yield uncertain
line-of-sight distances. Finally, in clusters, the high internal velocity dispersion will bias number 
densities towards lower values.   The distribution 
in environmental densities for our sample covers the range from -2.2 $\log$ \mpc \  to  1.23 $\log$ \mpc ~ with a median 
value of -0.67 $\log$ \mpc.  Our sample is underrepresented in the cluster regime; our 8 Virgo cluster galaxies sample the densest environment. The environmental density calculations were kindly provided by Jesse Miner.

\begin{figure}
\includegraphics[width=90mm]{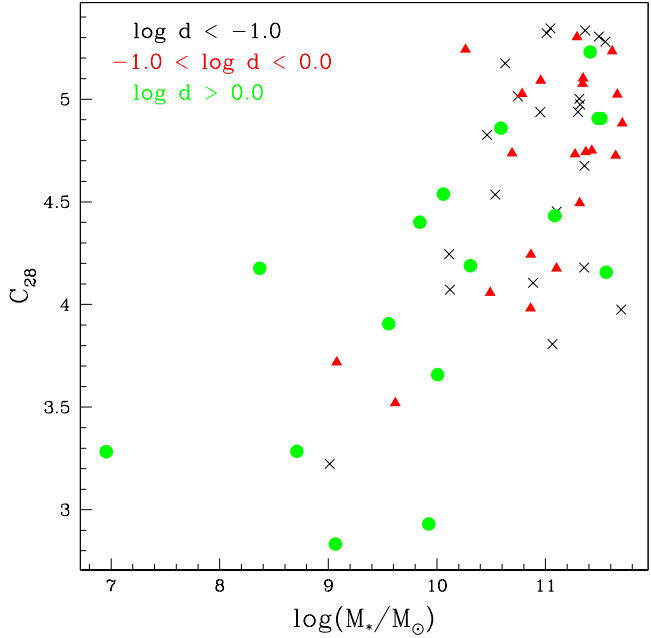}
\caption{Total stellar mass versus light concentration (C$_{28}$) for the sample of 59 galaxies. The point style designates local environmental density in \mpc (green circles: d $>$ 0.0,  red triangles: -1.0 $<$ d $\leq$ 0.0, and black crosses: d $\leq$ -1.0).}
\label{fig:massvsconc}
\end{figure}

Fig.~\ref{fig:massvsconc} demonstrates the broad parameter
space covered by our sample. Black crosses and green dots 
refer to galaxies in the lowest and highest density environments respectively. 
Although our S0 galaxies have mostly low concentration for low mass galaxies, SP trends in mass and concentration can still be identified.  However, the combination of a statistically incomplete sample regarding environment (e.g., all of the observed lowest mass galaxies are in the Virgo cluster high density environment) and a scarcity of high density environment galaxies makes it difficult to draw conclusions regarding the effect of environment on age or metallicity.   

\begin{figure}
\includegraphics[width=90mm]{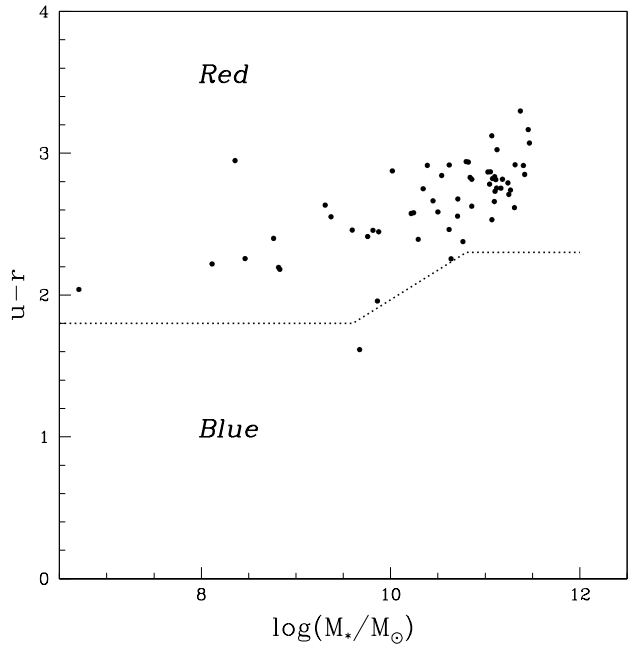}
\caption{Total color (u - $\emph{r}$) versus the stellar mass, in units of log solar mass. The dotted line from \citet{Kan08} separates the red and blue sequences of galaxies. Our full sample can be thought of as a collection of red sequence galaxies. }
\label{fig:umrvsmass}
\end{figure}

Galaxies typically occupy two distinct regions in color-stellar mass space; the so-called   ``red sequence'' and the ``blue cloud'' \citep{Str01, Kau03, Bel04,
Kan08}.
Although S0 galaxies generally fall into the red sequence, studies have shown 
that this morphology-color correspondence fails for low mass S0s 
\citep{Kan09}. Fig.~\ref{fig:umrvsmass} shows \emph{u}-\emph{r} color (Petrosian magnitudes from SDSS) 
versus stellar mass calculated using  \citet{Por04}.  The dotted line shows the
boundary between red and blue sequences, as reported by  \citet{Kan09} but with a mass offset factor of 1.8 to account for mass scale differences. We use the \citet{Por04} mass transformations to match with  \citet{Kan09}. All but one S0 galaxy (UGC09003) in our sample fall in the red sequence. 

\section{Stellar Population Models}\label{sect:stelpop}

A comparison of the observed galaxy colors with those predicted 
from SP synthesis models covering a range in age and
chemical composition allows for the determination of light-weighted
mean ages and metallicities. 
The combination of a primarily age sensitive color (e.g., \emph{g}-\emph{r}) with a 
primarily metallicity sensitive color (e.g., \emph{r}-H) provides a separation in age and 
metallicity. We use the \citet[][private communication, hereafter CB10]{CB09} simple 
stellar population (hereafter, SSP) model with a Salpeter initial mass function and STELIB libraries \citep{LeB03} for our analysis. Since only a single age is attached to any SSP model whilst
 the observed light likely results from multiple coeval populations,
 the derived ages of the stellar population contributing the integrated
 light are best understood as a light-weighted mean age rather than
 a mass-weighted age. From here on, we drop the ``mean'' and 
refer to these as light-weighted ages and metallicities to avoid confusion with mean 
ages that are radially averaged.

\begin{figure}
\includegraphics[width=90mm]{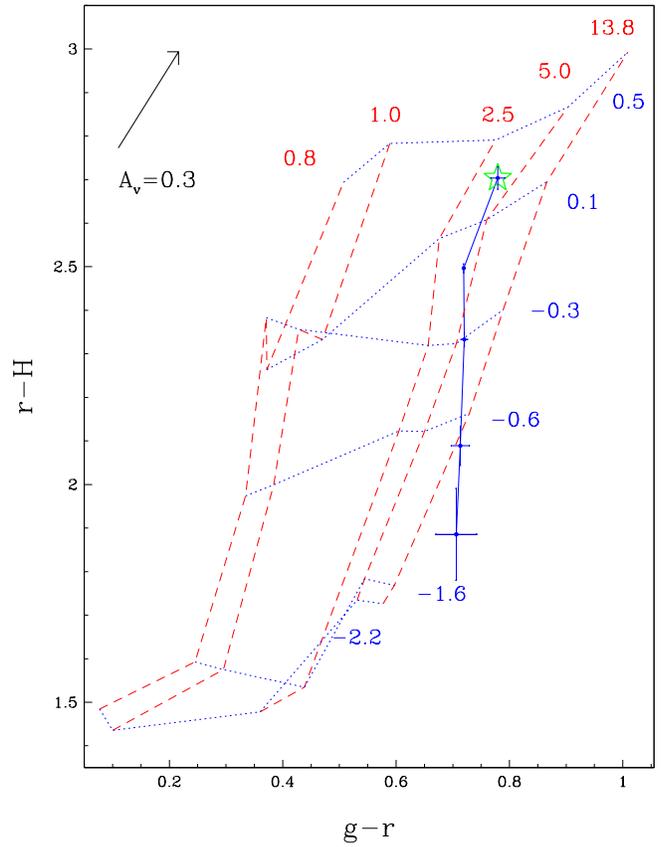}
\caption{ \emph{r}-H vs. \emph{g}-\emph{r} color-color diagram for UGC 10391. Galaxy colors are overlaid on a CB10 SSP model grid. Model lines of constant age (shown as dashed red lines) increase, left to right, from 0.8, 1.0, 2.5, 5.0, and 13.8 Gyr. Model lines of constant metallicity  (shown as dotted blue lines) increase from bottom to top from [Z/H] = -2.2, -1.6, -0.6, -0.3, 0.1, and 0.5. Small, filled circles are the average colors of the galaxy's radial bins (the binning scheme is noted in Table 5) and each bin is connected by the solid line. The central binning region is designated by a green star. The error bars on each radial bin represent $\pm$ 1 $\sigma$, where $\sigma$ is the standard error in the mean based on the scatter in color for each radial point in the designated bin added in quadrature to the sky error as discussed in section \S \ref{sect:sky}. A foreground screen dust model color vector with A$_v$ =  0.3 is shown in the upper left corner.}
\label{fig:grrh}
\end{figure}

Ages and metallicities are presented in Fig.~\ref{fig:grrh} for 
the UGC10391 color-color diagram. For each galaxy, the
central bin, subsequent radial bin, and consecutive radial bins are denoted by a green star, small filled circle, and a solid blue line, respectively. Overplotted is a CB10 SSP model grid. Red dashed lines 
represent model lines of constant age increasing, left to right, from 0.8 Gyr 
to 13.8 Gyr. Blue dotted lines represent model 
lines of constant metallicity increasing, bottom to top, from [Z/H] = -2.2 
to +0.5. The error bars for each radial bin represent the standard error in the mean based on the scatter in color of the radial 
points within the designated bin added in quadrature to the sky effect (see \S\ref{sect:sky}). This example shows a galaxy whose light-weighted
metallicity and light-weighted age decrease and increase respectively 
from the center of the galaxy outward. The galaxy gets bluer in 
\emph{r}-H at larger radii and keeps a roughly constant \emph{g}-\emph{r}, thus 
crossing over lines of constant age with increasing radius and indicating older 
ages in the outer regions of the galaxy. 

\subsection{Star Formation Histories and Model Uncertainties}\label{sect:sfh}

\begin{figure}
\includegraphics[width=90mm]{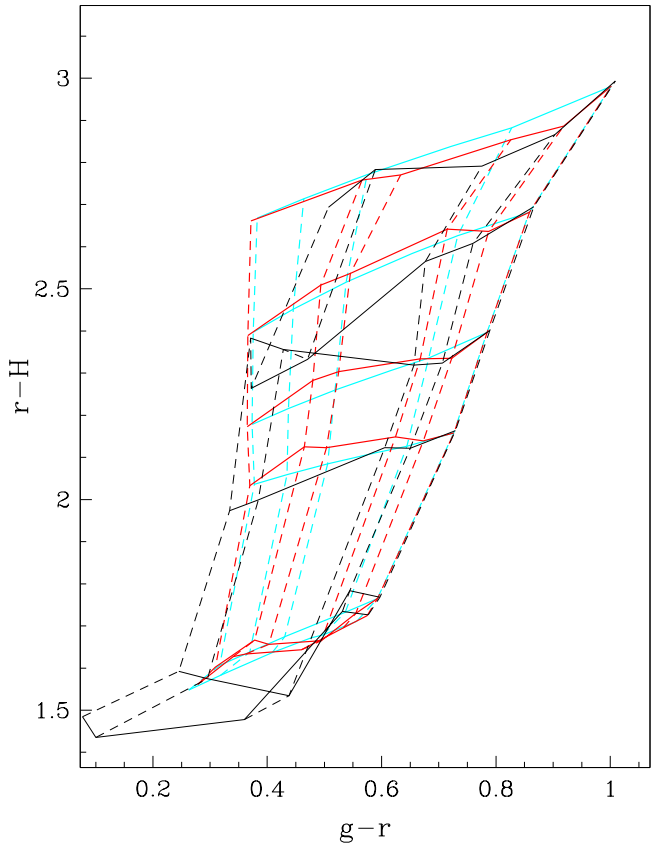}
\caption{ \emph{r}-H vs. \emph{g}-\emph{r} color-color diagram with CB10 models for various star formation histories. Overlaid are models using a constant star formation history (cyan), an exponentially declining star formation history (red), and a simple stellar population (black). Both the constant and exponential models assume a maximum age of 13 Gyr. Lines of constant age are dashed and lines of constant metallicity are solid. Metallicity increases from bottom to top for all 3 models as $[Z/H] = -2.2, -1.6, -0.6, -0.3, 0.1, 0.5$. Age lines for the SSP model are as in Fig.~\ref{fig:grrh}. Age lines for the exponential model increase from left to right as time constant $\tau =  100, 13, 6.5, 4.0, 3.0,$ and $0.1$. Age lines for the constant star formation model go from right to left as time since star formation has occurred $=$ 0.1, 10, 12, 12.8, 12.9, and 13 Gyr.}
\label{fig:modgrids}
\end{figure}

Since our analysis of SP trends in S0s hinges on the
reliability of 
derived ages and metallicities, we must 
understand the effects of SP 
model differences on our 
results. The SSP models that we use for our 
analysis predict the evolution in colors (and spectrum) 
of a coeval population of stars
with the same chemical composition and specified initial mass 
function. While this is clearly an over-simplification of actual star formation histories 
in S0 galaxies, it represents a straightforward way to obtain a 
{\it light-weighted} age and metallicity.
However, we must test how our analysis could be affected by the use of more
complex star 
formation histories. Since an SSP is one extreme star formation history 
(equivalent to a single burst), we compute models for the other extreme, a 
constant star formation history with a quenching of star formation at various 
ages, and compute models for the intermediate case of exponentially 
declining star formation. We thus consider populations that 
are composed of a superposition of SSPs, born at different epochs. Using the 
\emph{csp-galaxev} program\footnote{Part of the \citet{BC03} software release: http://www2.iap.fr/users/charlot/bc2003/}, we use an SSP 
model with a Salpeter initial mass function and fixed metallicity, convolved with the given star 
formation history (exponentially declining and constant). We compute the 
convolved models for a range of metallicities and time constants, $\tau$,  (for the 
exponentially declining star formation models) or star formation truncation times (for the 
constant star formation history models). In both cases, we set the age of the galaxy, or time 
that star formation began, to 13 Gyr. We show a color-color diagram 
based on the two sets of models in Fig.~\ref{fig:modgrids}. The constant star 
formation history is shown in blue, the exponentially declining star formation 
history in red, and the SSP model in black. The same SSP model grid is used throughout 
the paper. 
The meaning of the ``age'' of an SP for each set 
of models is different. For the exponentially declining models, we plot lines 
of constant $\tau$. For the constant star formation history models, the age is 
represented by truncation times, or how long the star formation has lasted since 13
Gyr. The model grid edges are similar between the three model sets. 
For example, the constant star formation model with a truncation time 
of 0.1 Gyr and the exponentially declining model with a $\tau$ of 0.1 Gyr 
are similar to 
the SSP model with an age of 13.8 Gyr, all lying and nearly overlapping near the right 
edge of the grids. The general shape of the lines do not vary with star formation 
history, except for the variation of the -0.3 metallicity line at young ages which is 
small compared to the large metallicity gradient that is observed in our sample. 
Thus, an analysis  based 
on relative age trends with radius will be robust with respect to star formation 
histories. 

Uncertainties in SP synthesis modeling have received notable attention \citep{Tra00,Sch02,Sha05,Con09}. 
Theoretical model uncertainties in the age and metallicity zeropoint can 
be caused by errors in the calibration of ages 
and metallicities from globular clusters. When two populations differ in metallicity as well 
as in age, and if the model zero point errors are metallicity dependent,
differential ages between metal-poor and metal-rich populations are less
secure. 
Indeed, there appears to be a strong metallicity gradient with 
radius for most of our galaxies. Therefore, our derived age trends 
with radius are subject to metallicity-dependent zero point issues and we take this into consideration when examining our results. Another concern is the difference in 
elemental abundance ratios between the population being studied and the stars 
from which the models are based. Massive S0 galaxies are known to have a 
higher abundance of alpha elements, relative to iron, as compared to solar 
neighborhood  stars \citep{Nel05,Tho05,Ser08,Bed08}. Thus when using stellar 
population models that do not account for non-solar abundance ratios, derived ages and 
metallicities are affected by errors in the theoretical stellar evolutionary tracks \citep{Sch07,Lee09}. 
Uncertainties in the input model
parameters, such as the effective temperature of the 
isochrones, giant stars or binary stars 
in the luminosity function, our understanding of late evolutionary phases, as well as the theory for convection and the effects of rotation 
and diffusion, may induce additional, unknown systematic errors. The treatment of advanced stages in evolution, such as the Thermal-Pulsating Asymptotic Giant Branch phase (TP-AGB), has received much attention lately. TP-AGB stars are extremely bright and dominate the NIR light of a galaxy following a burst of star formation, but represent a challenge for theoretical models due to  the combined effects of thermal pulses, changes from heavy element dredge-up, and mass loss \citep{BC03,Mar05}. 

The significant range in predictions between different SSP models can lead to great differences in the interpretation of galaxy colors. Although it is difficult to estimate uncertainties in the derived SP parameters (like age and metallicity), the comparison of available models suggests how reliable they are. The CB10 model is a revised version of the \citet[][hereafter, BC03]{BC03}  models, which includes a new prescription for TP-AGB evolution of low and intermediate mass stars  following \citet{Mar07}, and uses tracks and isochrones with updated input physics from \citet{Ber08}. \citet{Emi07} and \citet{Mac10} have highlighted significant changes with the new Bruzual and Charlot models in NIR model colors for intermediate populations. As well, the SP synthesis models of \citet{Mar05} different from the Bruzual and Charlot models by using a fuel-consumption-based code. These models place a particular emphasis on the TP-AGB phase and the effect on the model colors has been demonstrated in the literature \citep{Ton09}. 

\begin{figure}
\includegraphics[width=70mm]{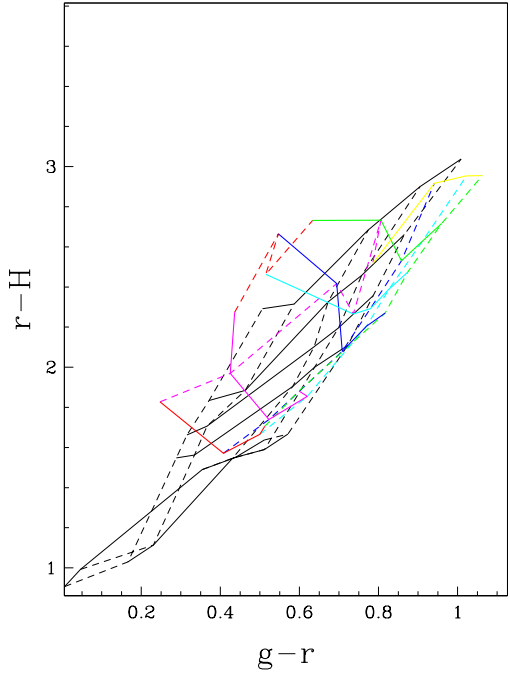}
\includegraphics[width=70mm]{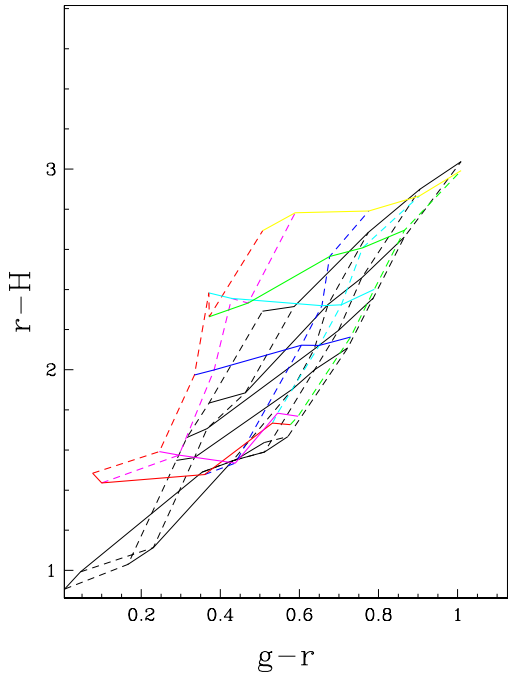}
\caption{Color-color diagrams using the BC03 model (in black) overlaid on other population synthesis models (in color). Lines of constant age and metallicity are represented by dashed and solid lines, respectively, and vary in color according to their values for all but the BC03 model. \citet{Mar05} and CB10 SSP model grids are shown in the top and bottom panels, respectively. \citet{Mar05} ages increase from left to right as 0.8, 2.0, 5.0, 10.0, and 14.0 Gyr SSP.  \citet{Mar05} metallicity increases from bottom to top as [Z/H] $=$ -2.25, -1.35, -0.33, 0.35, and 0.67 dex. The BC03 and  CB10 SSP model grids have ages and metallicities as in Fig.~\ref{fig:grrh}.}
\label{fig:marbc}
\end{figure}

 We compare the original BC03 SSP models to the SSP models of
\citet{Mar05} and CB10 in Fig.~\ref{fig:marbc}. \citet{Mar05} and CB10 SSP model grids are shown in the left and right panels, respectively. The BC03 model grid is shown in each panel (in black) for reference. All three models cover similar ages and metallicities (see figure caption for details), but the lowest age plotted here (0.8 Gyr) is not available for the lowest metallicity (-2.25 dex) in the \citet{Mar05} model. The CB10 models are a closer match to the \citet{Mar05} models than the earlier BC03 version. Both \citet{Mar05} and CB10 models are redder in \emph{r}-H, which provides a closer match to observations of star forming galaxies \citep{Emi07}.  Despite large differences in color at low ages, 
we find in all models that the lines of constant age are tilted in a similar direction at all ages. We will return to this key point in \S \ref{sect:ccplots}. For younger ages ($\simless$ 3 Gyr), the lines of constant metallicity in the CB10 and \citet{Mar05} models change slope (r-H colors decrease along lines of constant metallicity for increasing age) and lines of constant age vary slightly in their dependence on r-H colors.  We will discuss the model differences on colors in \S \ref{sect:results}.

\subsection{Extraction of Ages, Metallicities, and Gradients}\label{sect:extract}
 
For our analysis, we use the CB10 models with an SSP star formation history. We derive light-weighted ages and metallicities for each radial bin 
and for the inner and outer radial regions (as defined in \S\ref{sect:bin}) by fitting
CB10 SSP models to the galaxy colors. We compute a 
finely-spaced CB10 SSP model grid by interpolating linearly between the SSP 
metallicities and between the finely spaced ages provided by CB10. From the
\emph{g}-\emph{r} and \emph{r}-H (or \emph{r}-J) colors, we determine the ages and metallicities from the 
model that minimizes the difference between model and observed colors, calculated in 
quadrature. Due to the uncertainties in extrapolating model colors to 
larger metallicities and ages and the convergence in colors at 
large ages, the age and metallicity of a galaxy are set to a maximum of 13.8 Gyr and 
+0.5 dex, respectively. While the CB10 models extend to 20 Gyr, we chose a maximum age of 
13.8 Gyr for consistency with the current age of the universe \citep{Hin09} (the colors of 20 Gyr and 13.8 Gyr models are 
similar enough not to affect our analysis). 

The uncertainties in the derived ages and metallicities are determined via a Monte Carlo method based on M04. Two-hundred 
realizations of the model fits were performed for each radial bin and inner and outer regions
with the colors for each realization drawn from a Gaussian distribution of 
the errors in each color. The standard error in the mean is 
based on the scatter in color of the radial points within the designated bin or 
region added in quadrature to the sky effect. Ages and metallicities for each bin 
correspond to the mean of the ages and metallicities computed from each realization.
The positive and negative errors in the derived ages and metallicities for each bin are then taken 
as the interval containing 68\% of the 200 Monte Carlo realizations. If 
more than 5\% of the realizations produced either an age greater than 
13.8 Gyr or a metallicity greater than +0.5, we set the error to 
zero, indicating the lack of a measurable error.  
We note that model fitting 
errors do not reflect uncertainties in the model itself (see 
\S\ref{sect:sfh}). Ages, metallicities, and their uncertainties are listed in 
Table 3 for the inner and outer regions of all galaxies in 
our sample. Tables of ages and metallicities at each radial bin are available upon request.
 
\begin{figure}
\includegraphics[width=90mm]{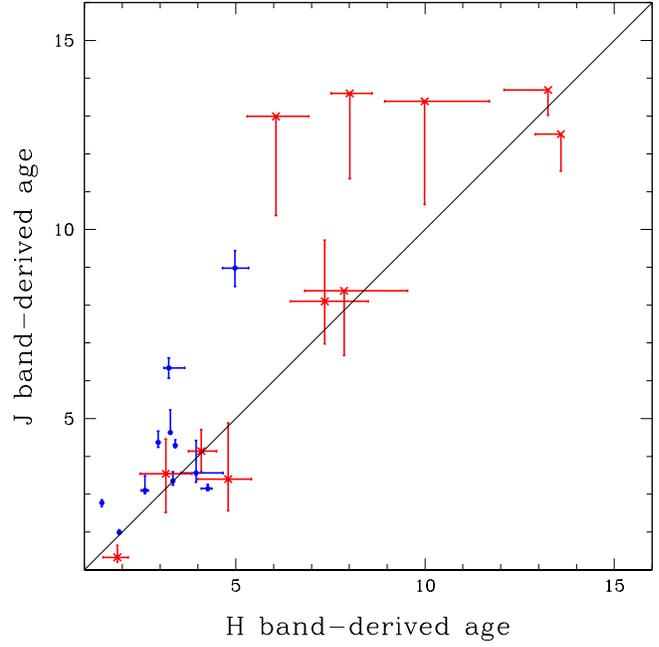}
\caption{Comparison of ages derived from J- and H-bands for the inner (blue circles) and outer radial regions (red crosses) for the 11 galaxies observed in both bands.  No error bar is shown when there is a lack of a measurable error. }
\label{fig:agecomp}
\end{figure}

There are 11 galaxies with available J and H-band images, enabling a partial consistency check on our extracted ages and metallicities. In Fig.~\ref{fig:agecomp}, we show the derived
J-band ages versus the derived H-band ages for both the inner and outer regions.
A Kolmogorov-Smirnov two-sample test (hereafter, KS) on ages for the 11
galaxies in common reveals a probability of 81\% that J and H-band derived ages are drawn from the same population for both inner and outer regions. 

Age and metallicity gradients are determined from a linear least-squares fit to
the age and metallicity data in the second, third, and fourth radial bins.  We avoid the
central radial bin due to the greater likelihood of dust contamination and seeing
blur; the fifth and sixth bins are avoided due to their greater errors from systematic sky 
uncertainties. The age and metallicity gradients for each galaxy are listed in the electronic version. Because the extracted ages and metallicities were set to a maximum of 13.8 Gyr and +0.5 dex
respectively, when an age (outermost bin) or metallicity (innermost bin) is set to one of those values, the gradient is considered a lower limit.

\section{Results}\label{sect:results}

We wish to constrain the star formation and chemical
enrichment histories of S0 galaxies by mapping light-weighted ages and
metallicities from the center out to large radii.  We first analyze the
central ages and metallicities.  While sky background removal is clearly the
limiting obstacle to accurate photometric colors at large radii, it is a
minimal problem in the central region.  In addition, there exists numerous age
determinations for the centers of early-type galaxies, including S0s, in the literature, thereby
enabling a useful comparison of modeling procedures.  On the other hand, central regions of early-type galaxies, which may include S0s, can be ``contaminated'' by recent star formation episodes that may be 
non-representative of the mean age of the rest of the galaxy 
\citep[e.g., ][]{DJD97,San07, Mac09}. Therefore the
mean age derived for the central region may depend on the 
aperture size for the age determination.  Moreover, PSF differences 
between different passbands make the interpretation of surface photometry within the
central regions of galaxies problematic.
Despite these complications, we first compare our central ages and metallicities  with corresponding values from the literature. An analysis of the galaxies' outer regions will follow. 
We note that our central regions all correspond to a radius of at 
least 3 arcseconds to alleviate PSF matching 
issues.

\subsection{Central Ages and Metallicities}\label{sect:centerage}

\begin{figure}
\includegraphics[width=90mm]{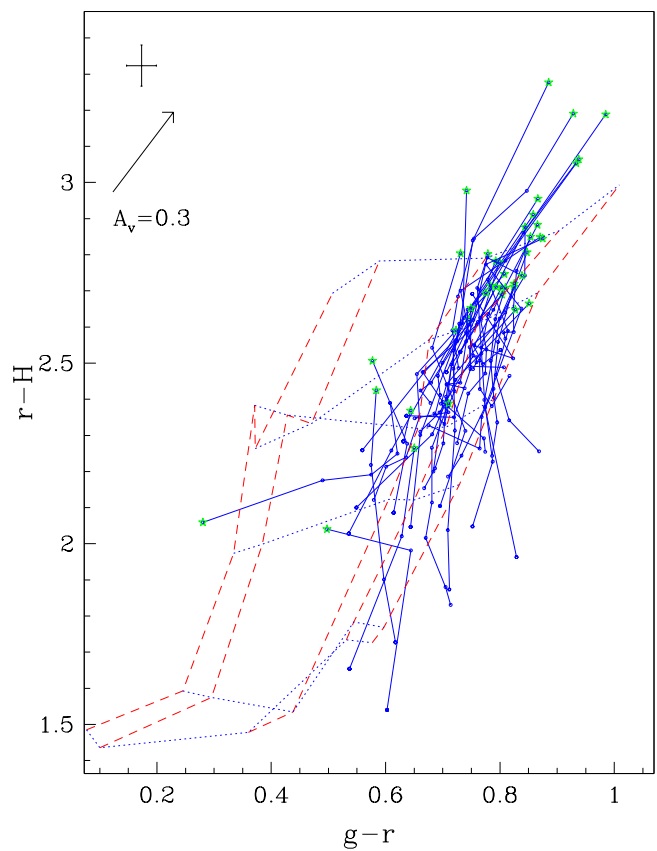}
\caption{Color-color diagram for the entire sample of 41 S0 galaxies with H-band imaging. The model grid is the same as in Fig.~\ref{fig:grrh}. The error bar in the upper left corner is the same as in Fig.~\ref{fig:grrh}, but  averaged for each galaxy at each radial bin. }
\label{fig:grrhallpsf}
\end{figure}

\begin{figure}
\includegraphics[width=90mm]{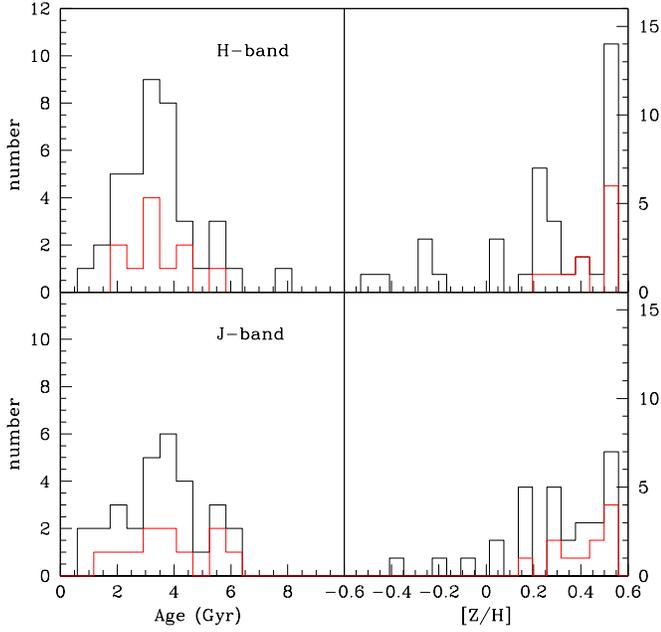}
\caption{Distribution of average age (left) and metallicity (right) of the central 0.5 R$_{e}$ for the full H-band sample, top, and J-band sample, bottom. Red lines show only galaxies that are observed in both the J- and H-bands.}
\label{fig:center}
\end{figure}

Central \emph{g}-\emph{r} colors versus \emph{r}-H colors are plotted in Fig.~\ref{fig:grrhallpsf} 
for all galaxies in the H-band 
sample relative to a SSP model grid, where the 
central bins (r $<$ 0.5 R$_{e}$) are 
designated by a green star. It can be seen that most of our galaxies 
have central colors that are concentrated in regions of intermediate age and
high metallicity. In 
Fig.~\ref{fig:center} we plot histograms of the
extracted central (r $<$ 0.5 R$_{e}$) ages and metallicities for the J- and H-band 
samples. The 11 galaxies with J- and H-band data  
yield similar age and metallicity histograms, therefore we shall merge the two data sets. 
However, we note non-uniformities in the J-band backgrounds that do not exist at H. 
Thus, when combining all the galaxies in our sample, the quoted ages and 
metallicities for galaxies that have H-band imaging will be based on that band alone.

The distribution of light-weighted central ages for the combined J- and H-band data is 
fairly symmetrical with a mean age of  3.6 $\pm$1.8 Gyr 
(the error is an rms scatter). The median central age is 3.3 Gyr.  
Thus, fairly recent star formation episodes have occurred
in the central regions of a majority of S0 galaxies. 
We can compare this result with similar studies 
of S0 galaxies based on: (1) optical and NIR photometry 
\citep{Pel99}, and (2) integral field unit (IFU) optical spectroscopy 
\citep{Sil06}, as well as comparing the ages and metallicities 
of specific galaxies in our sample with spectroscopically derived 
values. Since the central ages of early-type
galaxies, which may include S0s, are typically older for more massive galaxies \citep{CRC03,
Nel05,San06,Gra09,San07,Smi09}, we must  
compare galaxies of similar stellar mass (absolute luminosity). The typical stellar masses of 
our sample galaxies are similar to those of the 
Peletier and Sil'chenko samples. Thus, stellar 
masses cannot be responsible for any age 
differences between the samples.

From broadband
optical and NIR photometry, \citet{Pel99} found a mean bulge age of 9 $\pm$ 2 Gyr 
(at 1 bulge effective radius) for a 
sample of twenty S0 and early spiral galaxies (to Sbc). The ``bulges'' of
 \citet{Pel99} are closest in observed 
radius to our 2nd binning region, for which we find a mean 
age of 4.6 $\pm$ 2.4 Gyr.  While the
\citet{Pel99} ages are substantially greater than ours, these authors did not 
 include the seven bluest, youngest bulges in their sample, three of which are S0 
galaxies.   As well, they studied 
highly inclined galaxies and masked out the disk from their extracted ages and
metallicities, hence their results truly apply to bulge light at 1 R$_e$,
while ours apply to bulge and disk light mixed together within 0.5 to 
1.5 R$_e$. Finally, 
the difference in mean ages may be
partly attributed to the difference in SP models. \citet{Pel99} 
used the \citet{Vaz96} SSP models but note that the
\citet{Wor94} models give a significantly smaller mean age (2 Gyr) 
for their sample. The latter would be in closer agreement with our result. 

\citet{Sil06} compiled an IFU spectroscopic sample of 58 lenticular galaxies and used the Thomas et al (2003) 
models for age estimates. She reports median ages within the
unresolved nuclei of 3.7 and 6 Gyr for galaxies in sparse and
dense environments, respectively, while the median ages for `bulges' (extracted from an 
annulus of 4 to 7 arcsec) are 4.8 and 8.3 Gyr for sparse and dense
environments.  With the exception of the few Virgo galaxies, our sample shares the sparse environment of Sil'chenko's galaxies, and our central ages
are extracted from a region between the unresolved nucleus and
the 4-7 arcsec bulge.  Our age distribution
is thus a closer match to that of \citet{Sil06}.  

We also wish to perform a comparison of metallicities, for which the mean and median central  
values of our sample galaxies are [Z/H]= 0.2 $\pm$ 0.3 dex and 0.3 dex, respectively. 
The median value is preferable here, as it 
is more robust to the limit we place on metallicities at +0.5 dex. 
Our results are in
agreement with \citet{Sil06}, who finds a median metallicity for her nuclei 
and bulges of +0.4 dex and +0.2 dex, respectively. Visual inspection of
Fig. 3a in \citet{Pel99} reveals a mean metallicity near solar, 
consistent with our mean value of  [Z/H]=+0.0 $\pm$ 0.3 dex for our
second radial bin. 

Although our sample does not overlap with either \citet{Pel99} or \citet{Sil06}, 
independent ages and metallicities from spectroscopic studies offer direct 
comparisons for a few 
galaxies in our sample.  \citet{Kun06} extracted Lick index line strength measurements 
for UGC 5503. Comparison of  
their H$\beta$ and Fe5015 index measurements to a \citet{Vaz99} stellar 
population model yields an age of 1.6 Gyr for the inner 25". Our average 
extracted age for this region (comparable to 
our first two radial bins) is 1.3 Gyr. For UGC 10048, \citet{CRC03} estimated an age range 
of 4.9 to 7.9 Gyr, depending on the 
indices used. 
They used a 3\arcsc ~wide long slit spectrum with variance weighting,
thus it is difficult to compare specific regions in the galaxy. Nevertheless, our extracted age 
is slightly younger, ranging from 4.5 Gyr in the inner region to 3.7 Gyr in the 
outer region. The metallicity from \citet{CRC03} ranges from -0.1 to 0.1 while ours 
ranges from 0.1 in the inner region to 
-0.3  in the outer region. 
 \citet{CRC03} also studied VCC 1614, for which our extracted age for this galaxy ranges 
from 1.4 Gyr in the inner region to 5.5 Gyr in the outer 
region. This is consistent with their age of $\sim$1.8 Gyr. Our metallicity ranges 
from -0.4 in the inner region to -1.2 
in the outer region, consistent with the metallicity of $\sim$ -0.4 for \citet{CRC03}. 
In general, our 
extracted ages and metallicities are similar to those derived from spectroscopic studies. 

Fig.~\ref{fig:grrhallpsf} shows the center of several galaxies lying off the model grid toward larger \emph{r}-H values. 
The colors of these off-grid points may result from dust reddening. In many S0 galaxies, dust affects mainly the central regions \citep{Pel99, Wik01,Fri04}, although dust lanes are observed in 3-color optical images in the outer regions of a few galaxies in our sample. 
While our galaxy colors are corrected 
 for Galactic foreground extinction, internal extinction is ignored. 
To estimate the effect of internal extinction, a reddening vector for a foreground 
screen dust model 
with A$_v$ $=$ 0.3 is shown in the upper left corner of the color-color 
diagrams. Thus, if unaccounted for, intrinsic extinction biases our extracted
ages and metallicities towards higher values. This effect is clearly 
demonstrated and discussed in \citet{Pel99}. Their Figure 3a shows a large 
shift in colors, parallel to the reddening vectors, from the center of the 
galaxy to one R$_{e}$; ages could shift from $\sim$ 11 Gyr to $\sim$ 9 Gyr and 
metallicity from [Fe/H] $>$ 0.5 to $\sim$ 0.0. \citet{Pel99} found a dust signature 
in HST images in almost all S0 galaxies in their sample.

\subsection{Radial Age Trends}\label{sect:ccplots}

\begin{figure}
\includegraphics[width=90mm]{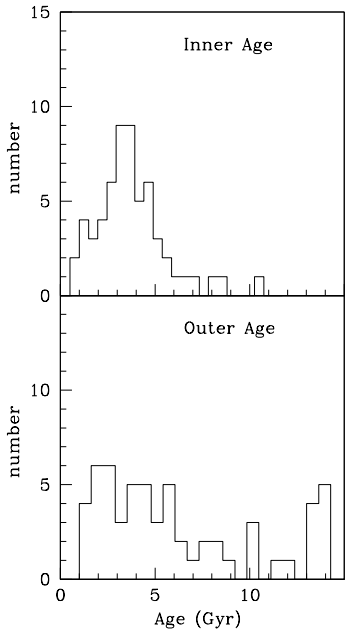}
\caption{Top and bottom panels show the distribution of average ages (in Gyr) of the inner (R $<$ 0.8 R$_{e}$) and outer  (R $>$ 1.2 R$_{e}$) regions, respectively, for all galaxies in our sample.}
\label{fig:outer}
\end{figure}

\begin{figure}
\includegraphics[width=90mm]{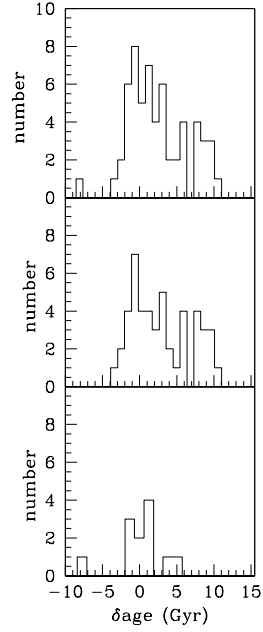}
\caption{Top: Distribution of light-weighted age differences (in Gyr) from the inner to outer regions (outer age - inner age) for all galaxies in our sample. Middle: Distribution of age differences for featureless disk galaxies. Bottom: Distribution of age differences for transition galaxies. }
\label{fig:agediff}
\end{figure}

While the inner regions of our S0 galaxies have light-weighted ages
of 3.7 $\pm$1.8 Gyr (the error is rms scatter), the outer regions are remarkably heterogeneous.  For those we find a mean age of 6.1 $\pm$ 4.0 Gyr. The $\pm$ 4.0 Gyr scatter in age exceeds the observational 
error. The larger age spread in the outer regions is notable by comparing the top and bottom panels of Fig.~\ref{fig:outer}, which shows all galaxies in our sample (as mentioned before, we use the 
H-band data for the 11 overlapping galaxies). Many galaxies have outer ages that exceed our limit of 13.8 Gyr. 
The increased heterogeneity 
in outer region ages compared to the inner regions naturally creates 
heterogeneous radial age trends, as illustrated in 
Fig.~\ref{fig:grrhallpsf}. We find an increase in age with 
radius from the inner to outer radial regions for 58\% of our sample of 59 galaxies, a decrease for
19\%, and a change of less than 1 Gyr for 23\%. 
In Fig.~\ref{fig:outer}, there is a clear 
tail in the outer region for high ages. As expected, this tail is also apparent in the 
distribution of age {\it differences} 
from the inner to the outer region, shown in the top panel of 
Fig.~\ref{fig:agediff}. Excluding this tail, many galaxies have similar ages  
between the inner and outer regions.

We find that the outer regions for 24\% of our galaxies show a substantial increase 
in light-weighted age from the center of the galaxy outward, resulting in outer 
regions older than 10 Gyr. We refer to these 
galaxies as ``old outskirts'' galaxies.  Specifically, we define an ``old 
outskirts'' galaxy to be one for which the integrated age of the ``outer'' 
region (i.e., all radii beyond $1.2~R_e$) is greater than 10 Gyr.  Note 
that some galaxies with ages $\less$ 10 Gyr in one or more outer radial bins 
are not classified as ``old outskirts'' galaxies because the integrated 
age for all radii beyond $1.2~R_e$ does not exceed 10 Gyr. However, we include two galaxies as ``old outkirts'' whose H-band derived ages in the outer region are just under 10 Gyr, but the J-band derived outer region ages exceeds 10 Gyr (see \S\ref{sect:oops}).
 Examination of optical color images for those galaxies shows no prominent features (i.e. bars, nascent spiral structure or rings) in their outer regions, except for UGC 9713 (bar) and UGC 10112 (faint ring). Color images for galaxies with old outskirts are shown in the electronic version and are indicated by an ``O'' above the image. 

We have veriÞed that the aforementioned results are robust to the
definition of the outer region by varying it from 1.2 to $3.5~R_{e}$. We
see no change and, thus, conclude that the definition of the outer
cutoff does not change strongly affect our results.  

We can compare our age trends with those from similar studies. Early photometric studies of S0 galaxies found mostly negative age 
gradients, but covered only the inner disk and used different color combinations than ours \citep{Cal83, BG90}.  \citet{Pel96} find an age difference
of less than 3 Gyr between the bulge and inner disk for most galaxies in their sample, similar to our peak at zero age difference. 
While they did not have many old outskirts galaxies, their study included early-type 
galaxies up to Sb, with only 8 pure S0 galaxies, two of which have older ages in the outer regions. 
Detailed studies of individual S0 galaxies have found 
both negative and positive age gradients 
\citep{Tik03,Nor06}. Finally, \citet{Sil06} found 
that, independent of environment, S0 galaxies contain a nuclear region that is 
younger than the bulge region. We show here that this trend continues 
outward for some galaxies in our sample, resulting in a large age gradient for 
those. 

We can compare our 
age gradients with those of  \citet{Fis96}, also discussed in \S \ref{sect:centerage},
and those of \citet{Raw09}, who studied a sample of 19 early-type galaxies, including 11 S0s, in the Shapley Supercluster core. 
For our galaxies, we find a mean 
$\Delta$age$/$$\Delta$log(r/R$_e$) $=$ 2.3 $\pm$ 4.6 Gyr dex$^{-1}$, with a median of 0.6 Gyr dex$^{-1}$. 
For a sample of 9 edge-on
 S0 galaxies, \citet{Fis96} found  
 $\Delta$age$/$$\Delta$log(r/R$_e$) $=$ 5.0 to 7.0 Gyr dex$^{-1}$ (depending on assumptions) for the bulge and 
 $\Delta$age$/$$\Delta$log(r/R$_e$) $<$ 1 Gyr dex$^{-1}$ for the inner disk. Even though we sample light from both the bulge and disk components due to the face-on nature of our 
 sample, we do find that our age 
 gradients are consistent. \citet{Raw09} found a lower mean age gradient of $\Delta$age$/$$\Delta$log(r/R$_e$) $ = $-0.10 $\pm$ 0.04 Gyr dex$^{-1}$, but also point out the existence of several individual galaxies with significant gradients in either direction, as we have found.

\subsubsection{Caveats for Old Outskirts S0 Galaxies}\label{sect:oops}

\begin{figure}
\includegraphics[width=90mm]{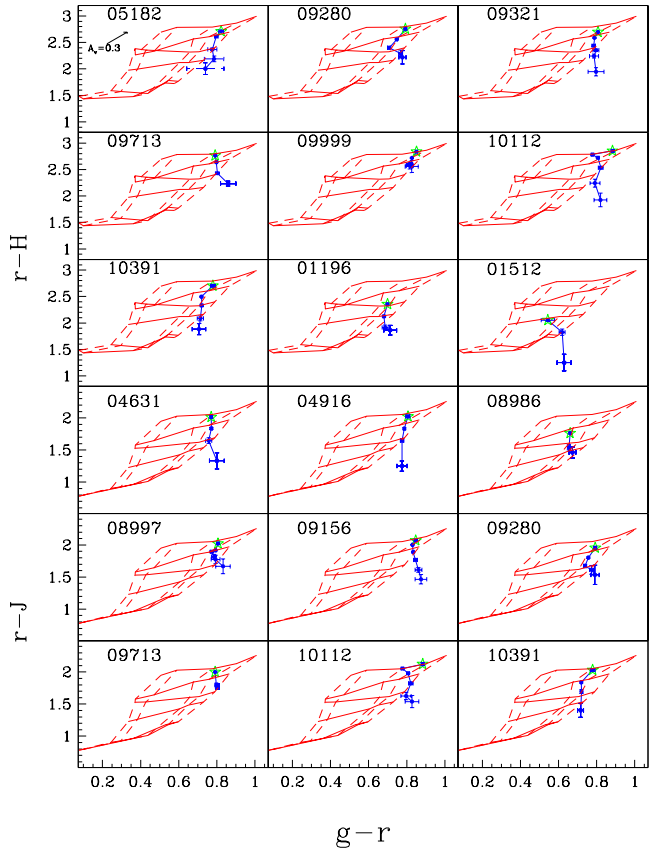}
\caption{Color-color diagrams for the sample of old outskirts galaxies in the H-band (top three rows) and J-band (bottom 3 rows) plot individually in a panel. The model grid and errors are the same as in Fig.~\ref{fig:grrh}. }
\label{fig:grrhsoo}
\end{figure}

A striking finding of this study is the significantly old ages in the outer regions of 14 S0 galaxies. We discuss these galaxies in more detail below along with potential caveats, whether observational or model dependent, for our inference of old outer regions. To assist our discussion, we show in Fig.~\ref{fig:grrhsoo} color-color diagrams for these 14 galaxies. 

We now test for observational consistency between J and H band data for the old outskirts galaxies. Based on H-band data, three galaxies, UGC 10163, UGC 10391 and UGC 9280, would not qualify as ``old outskirts'', but do so by virtue of J-band data. The H-band color-color diagrams for UGC 10391 and UGC 9280 show that those galaxies do have old outskirts, such as ages $>$ 10 Gyr in at least 2 outer radial bins (but narrowly fail to be classified as ``old outskirts'' galaxies based on their integrated ages beyond 1.2 R$_e$). Examination of the color image of UGC 10163 (see the electronic version) reveals dust in the outer region of the galaxy which artificially bias towards old ages. To be conservative, we eliminate this galaxy from our sample of old outskirts galaxies.  Two galaxies in our sample with J and H band data, UGC 9713 and UGC 10112, qualify as an old outskirts galaxy using ages extracted from both bands. Thus,  our final sample of old outskirts galaxies includes four S0s with data in both the J and H bands: UGC 9280, UGC 9713, UGC 10112, and UGC 10391. The color-color diagrams for those galaxies (Fig.~\ref{fig:grrhsoo}) are  very similar between J and H (the difference in UGC 9713 in the last bin is due to a stricter cutoff in the J-band).

Not only do old outskirts galaxies have old outer regions, they also have a positive age gradient.  Because of the tilt of the model grid in Fig.~\ref{fig:grrhsoo}, the outer region ages and age gradients are highly dependent on the \emph{r}-H color, even though the \emph{g}-\emph{r} color is more sensitive to age than \emph{r}-H. Thus, our results hinge on the observed large decrease in \emph{r}-H color. This underscores the importance of NIR colors in extracting ages from photometry.  

A major concern for the identification of faint outer regions in galaxies is the sky subtraction, which, as discussed in \S\ref{sect:profs}, is more problematic in the NIR than the optical. Possible systematic sky errors include a variation of sky level across the image field, a narrow field of view for the galaxy size (never a problem here), and a bias in the  sky level assessment. We discussed in \S\ref{sect:profs} NIR systematic error estimates  based on the variation in sky level from four boxes along the perimeter of the image, which is included in the error bars in Fig.~\ref{fig:grrhsoo}.  Provided our sky error estimates are correct, altering the sky background by a systematic error in most of these galaxies would not yield an age below 10 Gyr; that is, old outer regions remain. Finally, any error in our methodology in sky background estimates should affect all of our galaxies equally. 

It is also unlikely that the presence of dust has 
skewed our results of old outer regions. The reddening line shown in the top left of Fig.~\ref{fig:grrhsoo} shows the direction for dust reddening, which would make both $\emph{g}$-$\emph{r}$ and $\emph{r}$-H redder. Thus, a large amount of dust reddening would be necessary to explain the observed old ages. While S0 galaxies likely suffer some dust extinction,  \citet{Pel99} find from HST images that the light 
should be free from dust for most S0 galaxies beyond r $>$ $1~R_{e}$. Examination of SDSS color images of old outskirts galaxies shows no visible dust lanes or nascent spiral structure in the outer regions of these galaxies. 

We have also examined photometric zeropoint errors. However, because this would affect all radii equally, no amount of shift in any direction would negate the trend of increasing ages with radius for old outskirts galaxies.

Our derived ages are notoriously subject to modeling uncertainties, especially at NIR fluxes that are dominated by ill-constrained late evolutionary phases  \citep{Emi07, CB09,Mac10}. A major challenge for SP models is the treatment of the TP-AGB phase which affects mostly low ages and high metallicities, but old ages are unchanged.  However, colors for the outer regions of old outskirts galaxies fall off of all model grids, suggesting an inconsistency in observations and the models. While some of the offset may be due to sky subtraction error, this cannot explain the offset entirely. Thus it is possible that zero point errors in the SP models are at play and may affect our results. Models have been known to predict excessive galaxy ages \citep{Vaz01,Mac04,Tho05,Moo07}. Age dating of old stellar populations is influenced by uncertainties in the $T_{eff}$ and [Fe/H] scale of giant branch stars (a significant fraction of the continuum light of old stellar populations, even in the blue, is provided  by giant stars) as well as the luminosity function of the upper red giant branch \citep[][hereafter S02]{Sch02}. For example, S02 found that the unrealistically large derived age for the globular cluster 47 Tucanae from comparison of its observed integrated spectrum to population synthesis models may be partially explained by the discrepancy between the observed luminosity 
function of the upper red giant branch of the cluster and the lower expected number based on synthesis models. Indeed, the model predictions in S02 for (B-V) are too blue compared to the observations of 47 Tucanae, so that an older age is needed to reconcile with the observed color. Despite these concerns for the model-dependent derivation of old ages, it is unlikely that the result of old outer regions in a sub-sample of S0 galaxies will be revised with improved models.

Finally, our reliance on SSP models to extract ages should not affect this result. Since young stars dominate the light, an SSP age greater than 10 Gyr indicates the lack of any significant star formation for quite some time. Therefore, no star formation history is necessary.  

Given all of the above considerations, we feel that the result of old galaxy outskirts is robust in most cases.

\subsection{Age Trends With Galaxy Properties}\label{sect:agetrend}

\begin{figure}
\includegraphics[width=70mm]{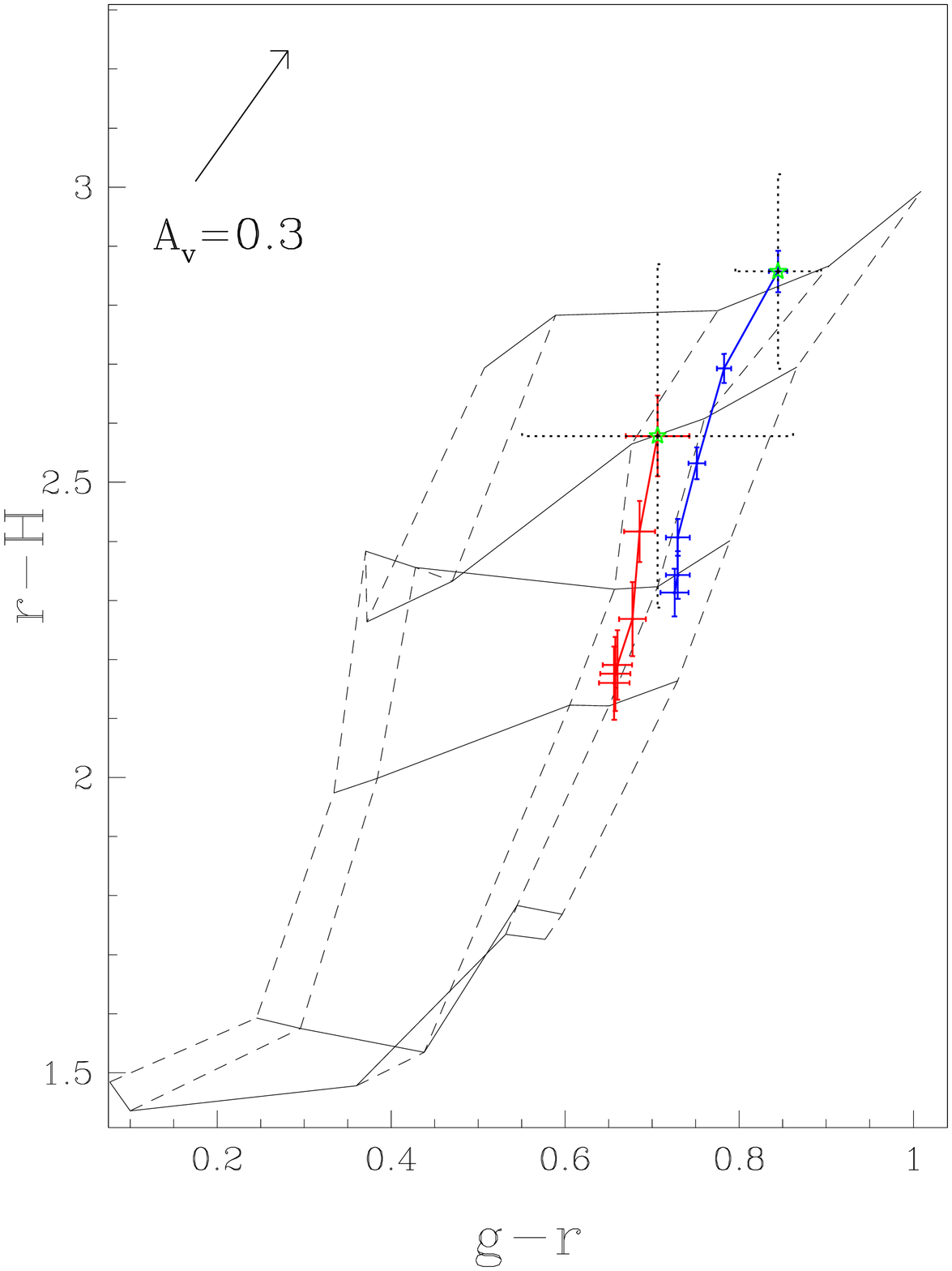}
\includegraphics[width=70mm]{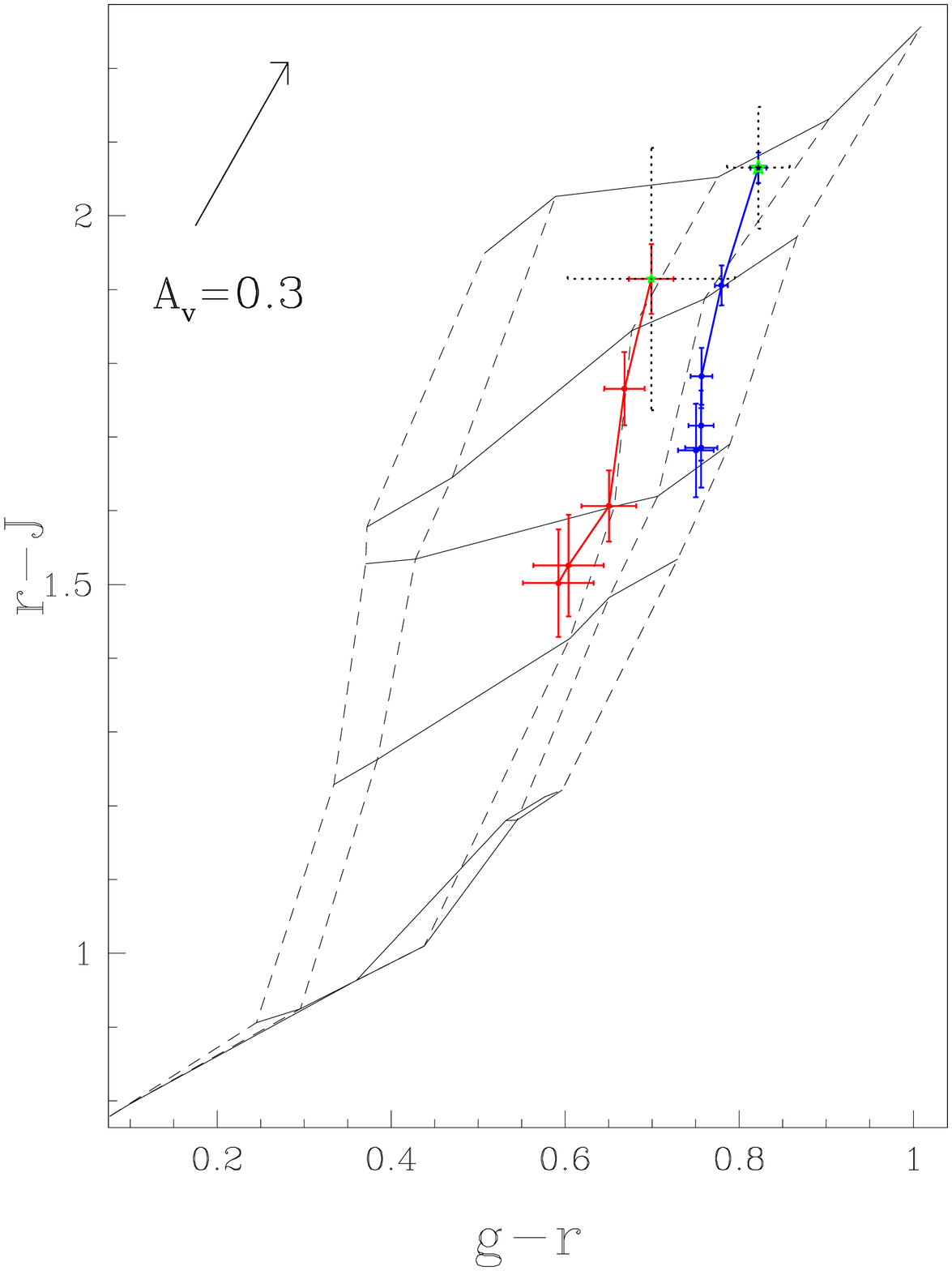}
\caption{Color-color diagrams for the low (red) and high (blue) stellar mass groups. The same model grid is used as in Fig.~\ref{fig:grrh}; the top panel shows \emph{r}-H vs \emph{g}-\emph{r} for the H-band sample and the bottom panel shows \emph{r}-J vs \emph{g}-\emph{r} for the J-band sample. The red and blue solid lines delineate galaxies with mass $<$ $1\times 10^{11}$ $M_{\sun}$ and mass $>$ $1\times 10^{11}$ $M_{\sun}$, respectively.  The solid colored error bars on each radial bin represent the standard error in the mean based on the scatter in color among galaxies combined in quadrature with the standard error in the mean of the systematic sky error. The black dotted error bar on the first radial bin represents the galaxy-to-galaxy scatter. The reddening line, ages and metallicities, and symbols are the same as in Fig.~\ref{fig:grrh}. }
\label{fig:grrhsortonlymass}
\end{figure}

We have established that S0 galaxies exhibit a variety
of radial trends in age, with a substantial population having large positive age gradients. 
We now inquire whether these age trends, as well as the mean ages, 
are correlated with global properties such as total mass and light concentration.  To that end, we have separated our sample of 59
S0 galaxies into two roughly equal groups of low and high mass galaxies, straddling 1$\times10^{11} $ M$_{\sun}$.  The colors in each radial
bin are averaged for all galaxies in each group and are plotted in the \emph{g}-\emph{r} vs \emph{r}-H diagram (left panel) and \emph{g}-\emph{r} vs \emph{r}-J diagram 
(right panel) in Fig.~\ref{fig:grrhsortonlymass}.  The solid colored error bars 
are the standard error in the mean for each radial bin. The black dotted error bars 
in the first radial bin denote the 1$\sigma$ scatter in color for galaxies in that bin. 
In both panels a separation in ages is apparent between the two mass groups, such that the stellar 
populations of less massive galaxies 
are younger on average than those of more massive galaxies at all radial bins; this effect is more pronounced in the J-band sample.  
Although the sample is not statistically complete, accounting for possible environment-age trends \citep[as observed in][]{Sil06} would only strengthen our mass-age relation. A KS
two-sample test was applied to the high and low mass age distributions for all radial bins combined for all galaxies in our sample.
The hypothesis that the two samples are drawn from the same population in mean age can be rejected, with a p-value of 1.7$\times$10$^{-5}$, confirming what is apparent in Fig.~\ref{fig:grrhsortonlymass}.  The results of this
and other KS tests are summarized in Table 7.  
When only the inner regions are considered, the ages of the two mass groups can still be rejected as a single population (at the 97\% significance level).  This result is
consistent with \citet{Sil06}, who found that the stellar populations of more massive S0 bulges 
are older than those of less massive bulges. However, the bias in mass and environment of our sample (see \S\ref{sect:param}) may weaken age-mass relations. 

\begin{figure}
\includegraphics[width=70mm]{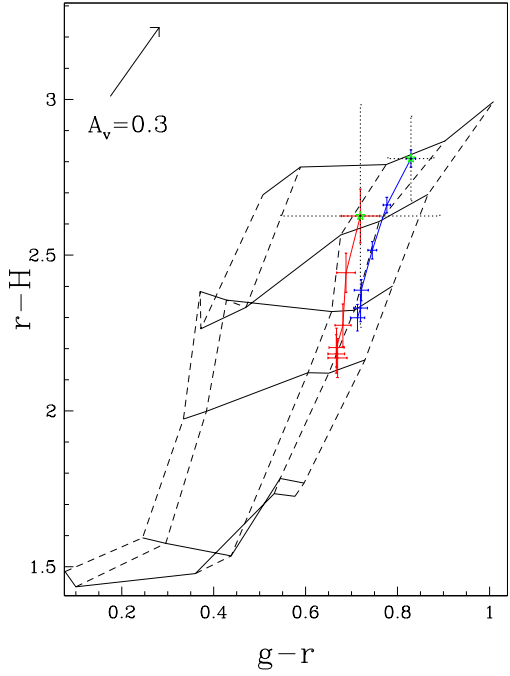}
\includegraphics[width=70mm]{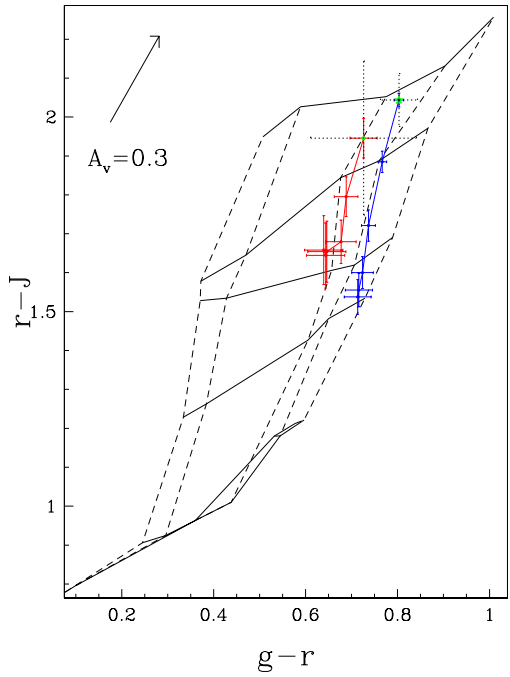}
\caption{Color-color diagrams for galaxies separated by high and low light concentration. The solid red and blue lines represent the average colors for galaxies with $C_{28}$ $ <$  4.7 and $C_{28}$ $>$ 4.7, respectively. The model grid and error bars are as in Fig.~\ref{fig:grrhsortonlymass}.}
\label{fig:grrhsortonlycon}
\end{figure}

\begin{figure}
\includegraphics[width=70mm]{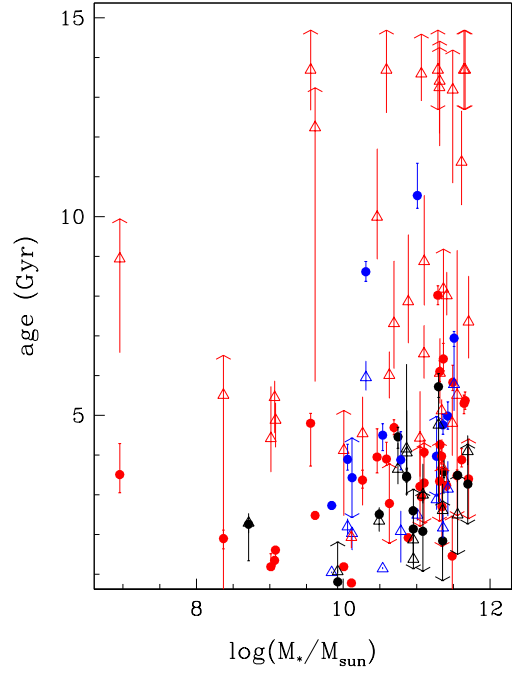}
\includegraphics[width=70mm]{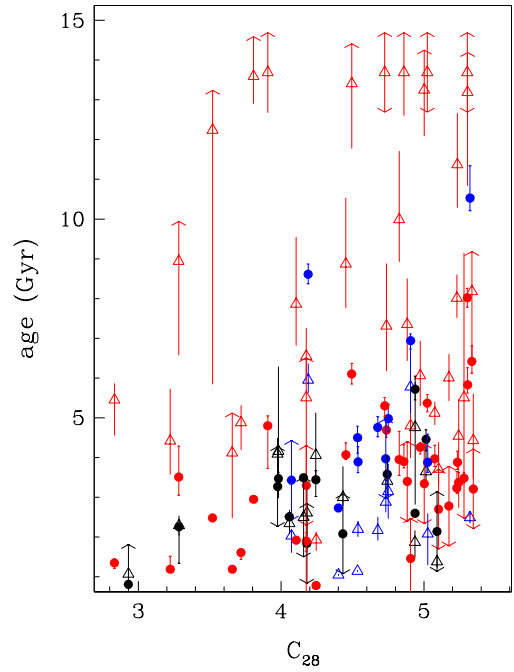}
\caption{Age versus stellar mass (top panel) and concentration (bottom panel) for inner (solid circle) and outer (open triangle) radial regions. Colors denote the radial trend; red indicates an increase in age from the inner to the outer region, blue indicates a decrease in age, and black indicates $\Delta$Age $<$ 1 Gyr. Arrows on the error bars indicate the lack of a measurable error.}
\label{fig:agemassconc}
\end{figure}

We have also sorted galaxies into high 
and low concentration groups, using a separation at the median value of the 
concentration index, $C_{28} = 4.7$. As 
above, the ages at each radial bin were combined for each galaxy in the two
concentration subsamples. Fig.~\ref{fig:grrhsortonlycon} shows a 
color-color diagram for the 
J- and H-band samples with the average colors of each concentration group averaged.  A  separation
in age is evident, such that the stellar populations of centrally concentrated galaxies 
($C_{28} > 4.7$)
are older than those of less centrally concentrated galaxies. As with mass, the J-band sample, which does not include the Virgo galaxies, shows a clearer separation in age. The hypothesis of a similar parent distribution in age between the two concentration groups can be rejected, with a p-value of 1.9$\times$10$^{-4}$, when all radial bins are combined. We do not find a statistically significant (within 1 $\sigma$) age difference in the outer regions alone for different concentration groups. To show  the mass-age and concentration-age relations without the constriction of groups, we plot in Fig.~\ref{fig:agemassconc} age versus mass and concentration for the inner and outer radial regions.

\begin{figure}
\includegraphics[width=70mm]{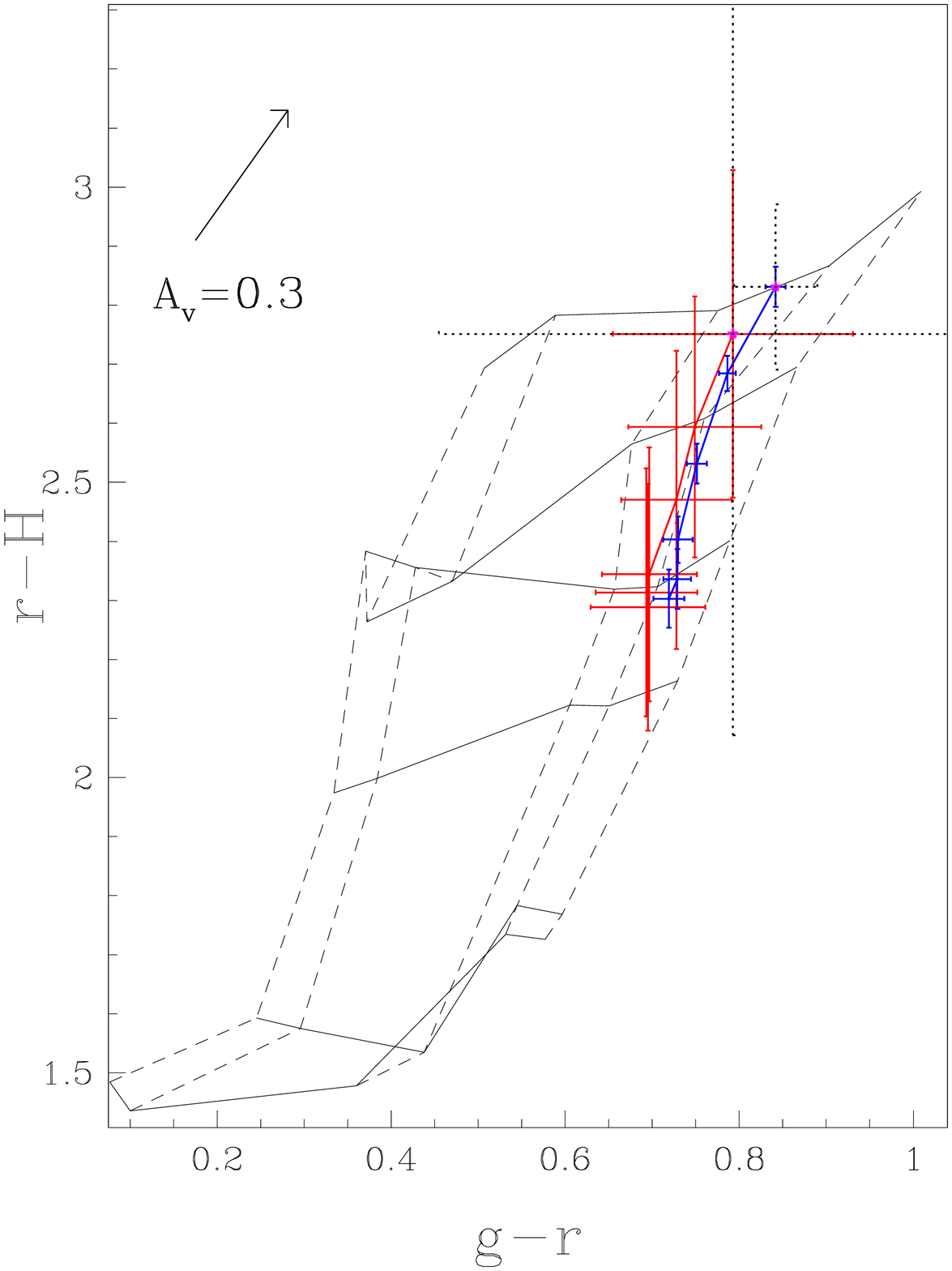}
\includegraphics[width=70mm]{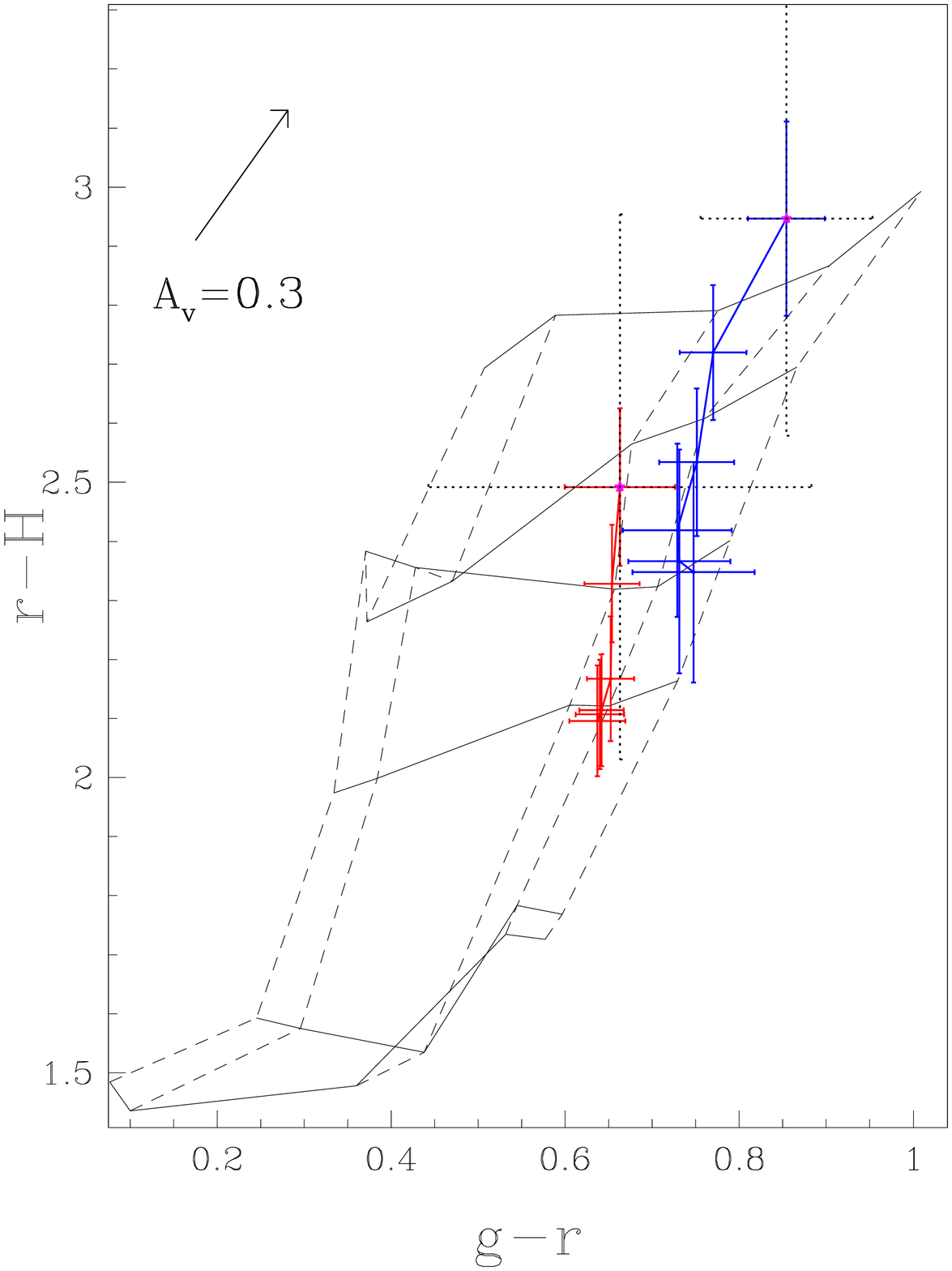}
\caption{Color-color diagrams for high (top) and low (bottom) light concentration galaxies, separated by $C_{28}$ $=$ 4.7 are shown for the H-band sample. The solid red and blue line represents the average colors for galaxies with mass $<$ $1\times10^{11}$ $M_{\sun}$ and mass $>$ $1\times10^{11}$ $M_{\sun}$ , respectively.  The model grid and error bars are as in Fig.~\ref{fig:grrhsortonlymass}.}
\label{fig:grrjsortsepmass}
\end{figure}

\begin{figure}
\includegraphics[width=70mm]{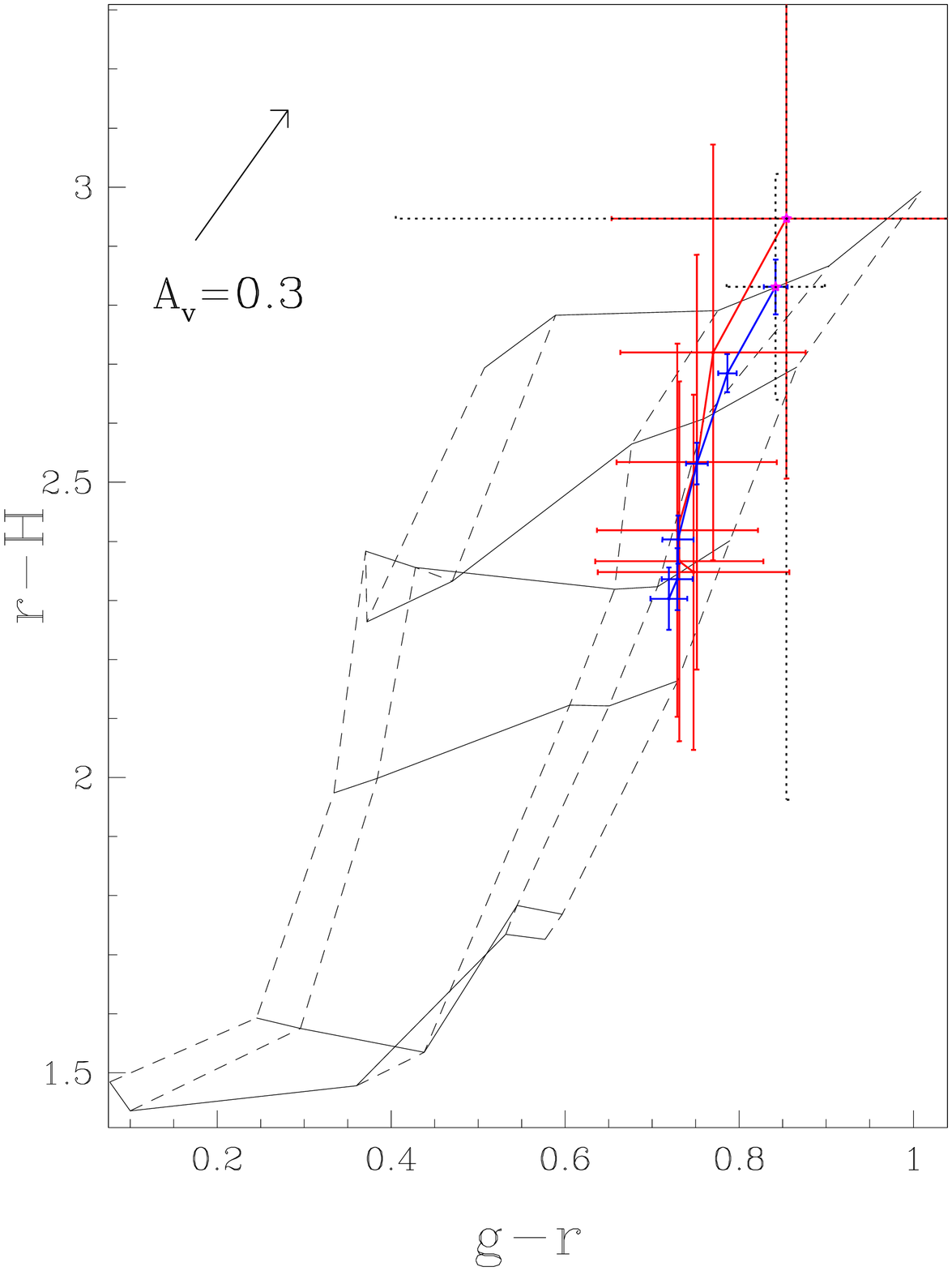}
\includegraphics[width=70mm]{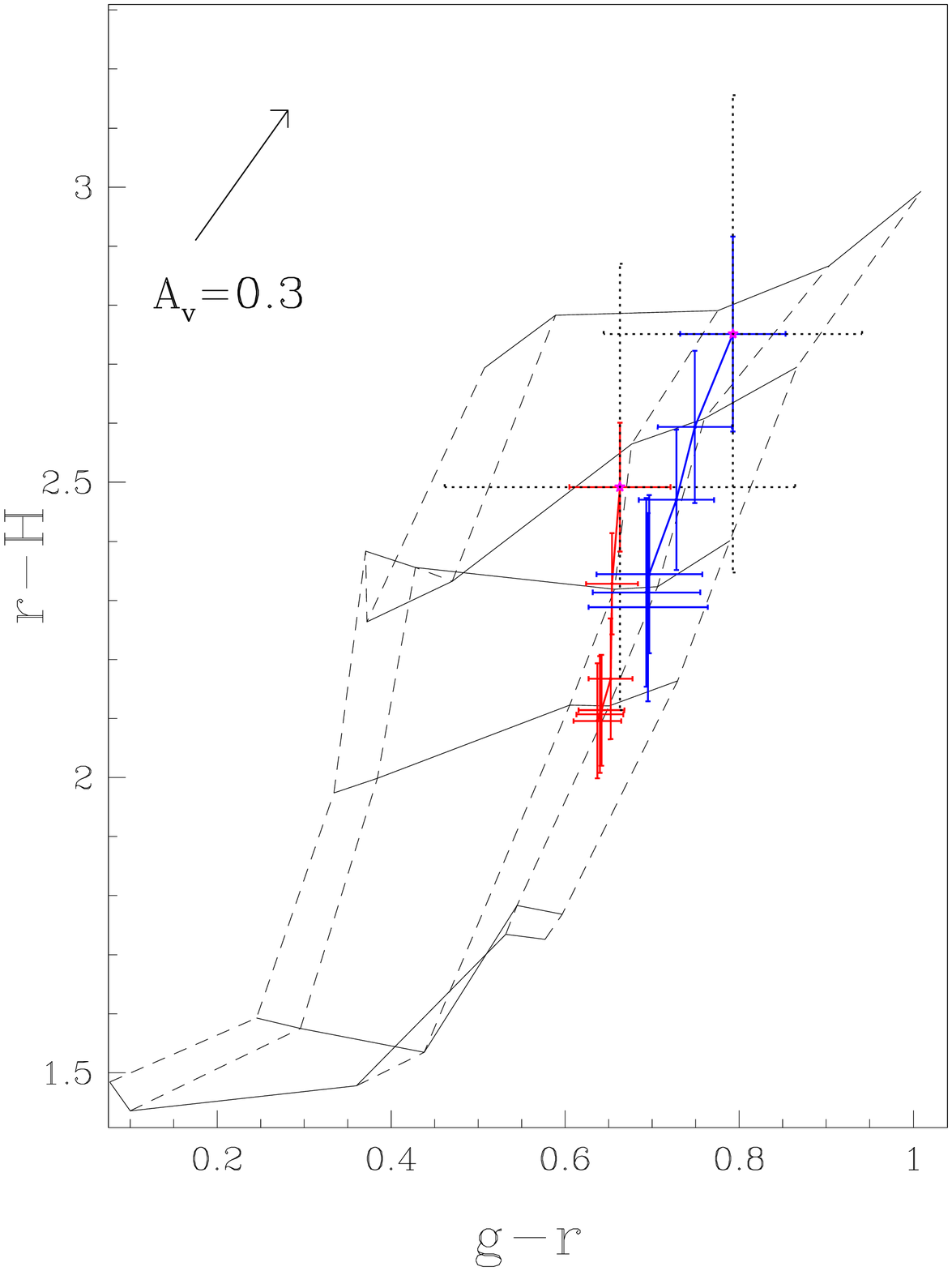}
\caption{Color-color diagrams for high (top) and low (bottom) mass galaxies, separated by a mass of $1\times10^{11}$ $M_{\sun}$, are shown for the H-band sample. The solid red and blue line represents the average colors for galaxies with $C_{28}$ $<$ 4.7  and $C_{28}$ $>$ 4.7, respectively.  The model grid and error bars are as in Fig.~\ref{fig:grrhsortonlymass}.}
\label{fig:grrjsortsepconc}
\end{figure}

Because low concentration galaxies tend to be less massive than higher concentration 
ones, as seen in Fig.~\ref{fig:massvsconc}, the age trends found for both
mass and concentration could be intertwined, with only one of the variables
being the driving factor.  
To test which  of mass or concentration is more fundamental, we further subdivide our
high mass sample into low and high concentration subsamples, and do the same for
the low mass sample. 
In Fig.~\ref{fig:grrjsortsepmass} color-color diagrams are plotted with  
galaxies separated by mass as in Fig.~\ref{fig:grrhsortonlymass}, 
however this time only high concentration galaxies are plotted in the left 
panel and only low concentration galaxies are plotted in the right panel. Fig.~\ref{fig:grrjsortsepconc} reverses the analysis. Figs.~\ref{fig:grrjsortsepmass} - \ref{fig:grrjsortsepconc} show a larger separation in mean age for the low concentration and low mass galaxies. While this may suggest that both parameters drive the correlation in age, these eyeball tests are not necessarily statistically sound (at least within 1 $\sigma$). Combining mass and concentration, we find that galaxies with both low mass and low concentration are younger in mean age than the 
rest of our sample in the inner region (2.9 Gyr versus 4.2 Gyr).

Our sample of S0 galaxies is a morphological mix, with some 
galaxies showing more spiral features than others. We now examine 
how radial age trends are 
dependent on morphology.  
The distributions of age differences for non-transition galaxies 
(featureless disk) and for transition galaxies are plotted in Fig.~\ref{fig:agediff} in the middle and 
bottom panels respectively. There is a positive age gradient for featureless disk S0s which is
clearly stronger than that for transition galaxies. 
Transition galaxies do not appear to have
the same tail of large age 
differences as other galaxies. This is expected if  
transition galaxies have a growing spiral structure where young stellar 
populations are located. 
A KS test for the age differences of transition and non-transition groups reveals statistically different populations at a greater than 3$\sigma$ confidence level. 

We note that old outksirts galaxies do not exhibit a preference for any particular mass or environment, but may for concentration.  Also, age $\emph{gradients}$, do not demonstrate a correlation with mass or concentration. 
 
\subsection{Metallicity Trends}

Negative metallicity gradients have been previously found for S0 
galaxies \citep{Fis96, Tam03, Ric09,Raw09}. From their spectroscopic sample 
\citet{Fis96} found a mean of $\Delta$[Fe/H]$/$$\Delta$log(r/R$_e$) $=$ -0.9 to -0.7 
(depending on assumptions) for their bulges and 
$\Delta$[Fe/H]$/$$\Delta$(r$_{kpc}$) $=$ -0.04 to -0.06 kpc$^{-1}$ for the disks. \citet{Raw09} found a mean metallicity gradient of $\Delta$[Z/H]$/$$\Delta$log(r/R$_e$) $=$ -0.13 $\pm$ 0.04. 
We carry our measurements typically 2.5 to 5 times further in radius and look for correlations with global 
properties. 
Since colors alone do not yield information about non-solar element abundance ratios, 
we examine the trends of mean metallicity in S0 galaxies based on the scaled
solar modeling described in \S 6.  

Fig.~\ref{fig:grrhallpsf} shows that 
virtually all S0s have significant negative metallicity gradients. 
We find a mean metallicity gradient $\Delta[Z/H]/\Delta log(r)$ $=$ -0.6 $\pm$ 0.5. In terms of physical radius,
the gradients have a mean $\Delta[Z/H]/\Delta(r_{kpc})$ 
$=$ -0.1 $\pm$ 0.3, consistent within the errors with the above studies.  Recall that the central-most binning region is avoided in the calculation of gradients.  This largely alleviates any artificial metallicity bias due to dust, if any, in galaxy centers. 

The correlation of metallicity with mass and concentration 
can be assessed in the same way as for age, i.e., by dividing the 
sample into high and low mass and concentration groups. For the trends of inner region metallicities, eleven galaxies lie off the model grid and are thus set to the maximum 
metallicity of 0.5 dex; this would most likely have a weakening effect on metallicity trends. As well, accounting for the fact that all of the observed lowest mass and lowest concentration galaxies are in the Virgo cluster (high density environment) would only strengthen the mass and concentration trends that we observe, because galaxies in higher density environments are found to have higher metallicity (at a given mass) \citep{Coo08}. 

\begin{figure}
\includegraphics[width=70mm]{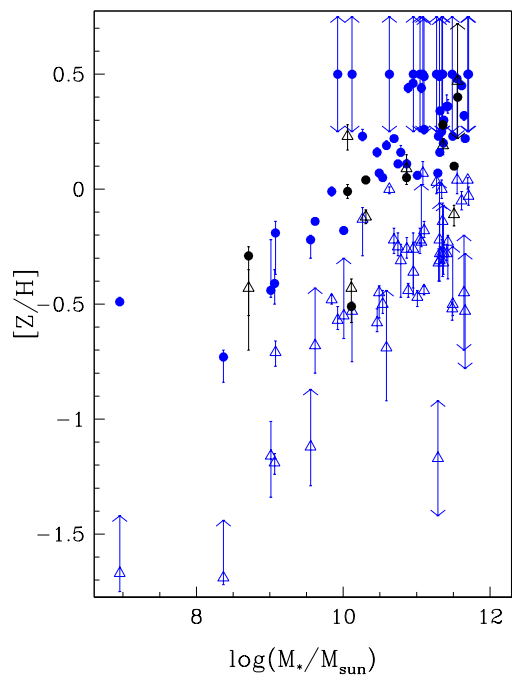}
\includegraphics[width=70mm]{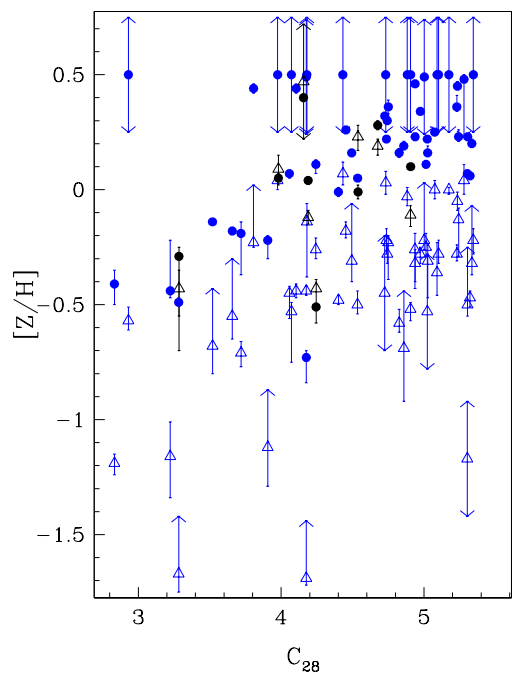}
\caption{Metallicity versus stellar mass (top panel) and concentration (bottom panel) for inner (solid circle) and outer (open triangle) radial regions. Colors denote the radial trend; blue indicates a decrease in metallicity from the inner to the outer region and black indicates $\Delta$ [Z/H] $<$ 0.25. Arrows on the error bars indicate the lack of a measurable error.}
\label{fig:metmassconc}
\end{figure}

Examination of Figs. \ref{fig:grrhsortonlymass} and \ref{fig:grrhsortonlycon} shows a strong separation in metallicity for the high versus low mass and concentration groups such that low mass or low concentration galaxies have lower metallicities. 
High and low mass galaxies have statistically different metallicity distributions in both the inner and outer regions (KS test reveals a p-value of 6.77$\times$10$^{-5}$ for the inner regions and 0.01 for the outer regions).  Studies of mass-metallicity relations in S0 galaxies are abundant \citep{Gar02,Kau03,Mac04,Tre04,Tho05}, but most of these studies report a central metallicity ($<$ 1 disk scale length) or a global metallicity (such as studies at high z where the aperture encompasses the entire galaxy) so that a comprehensive mass-metallicity relation in the outer parts of S0 galaxies is still missing.  The metallicity-concentration trends we see may be related to the increase in metallicity with early Hubble type (M04).  Metallicity $\emph{gradients}$, on the other hand, do not demonstrate a correlation with mass or concentration. The mass-metallicity and concentration-metallicity relations are also demonstrated  in Fig.~\ref{fig:metmassconc} for the inner and outer radial regions.

We try again to separate the effects of mass and concentration on metallicity. Examination of Figs. ~\ref{fig:grrjsortsepmass} and \ref{fig:grrjsortsepconc} shows a significantly larger separation in mean metallicity for the low concentration and low mass galaxies. 
For the inner regions, the hypothesis that high and low mass galaxies of low concentration are from the same population and that high and low concentration galaxies of low mass are from the same population can be rejected, with a p-value of 2.5$\times$ 10$^{-4}$ for both. There is no statistical difference for the high mass galaxies. That the high mass sample does not show a difference in metallicity for concentration groups may either mean that concentration does not play a large role in metallicity trends or that it does not play a role in the range of concentration covered ($C_{28} > 3.8$). Regardless of the driving force for the metallicity relations,  we find that galaxies with both low mass and low concentration have a lower mean light-weighted metallicity in the inner regions than other galaxies in our sample.

\section{Discussion}\label{sect:discussion}

We have examined the connection between the globally averaged light-weighted ages and metallicities 
of S0 galaxies and other global properties, such as stellar mass and concentration. We have also investigated age and metallicity gradients 
in S0s. We find that at all radii, S0 
galaxies with lower mass and lower concentration have younger light-weighted ages and lower light-weighted metallicities. 
We have also found that virtually all S0 galaxies in our sample  
have negative metallicity gradients, with an average value of
$\Delta$[Z/H]$/$$\Delta$log(r/R$_e$) $=$ -0.6 $\pm$ 0.5, consistent with published results.  On the other hand, the
radial behavior of age in S0s is heterogeneous; both positive and negative age 
gradients are found.
We observe an increase in age
with radius for 58\% of our sample, a decrease for 19\%, and 
little change ($<$ 1 Gyr) for 23\%, and a mean age gradient of $\Delta$age$/$$\Delta$log(r/R$_e$) $=$ 
2.3 $\pm$ 4.6 Gyr dex$^{-1}$. For 24\% of our sample, we find populations with 
substantially old light-weighted ages ($>$ 10 Gyr) in the outer region of the galaxy. 
 We now consider the implications of these results.  

\subsection{Mass and Concentration Relations}

The mass-metallicity trends that we observe extend into the outer regions of the galaxy, which may be dominated by a disk component, but we can not be certain without full bulge-disk decompositions of our galaxies. We assume here that the inner regions are bulge dominated, but this, as well, is not certain without decomposition analysis. 

Both high galaxy concentration and high galaxy mass point to a higher bulge mass. We speculate whether a relation between the bulge mass with age and metallicity in the inner and outer regions favor a particular model for S0 formation. 
Observations consistent with bulge growth through internal (i.e. driven by disk instabilities) and external (i.e. driven by satellites) secular evolution have been noted in various studies of late type spiral galaxies \citep{Cou96b,Com00,Ell01,Kan04,Kor04,Barw07}. A small 
bulge, such as in late type spirals or low mass, low concentration S0s, is more likely to contain associated star formation from 
secular evolution, even if most of the mass has an earlier origin \citep{Ell01}. 
Indeed, the spectroscopic analysis of stellar populations by \citet{Mac09} reveals 
that $\sim$ 20\% of the mass of late type spiral bulges consists
of stars less than 1Gyr (although, 70\% of the bulge $\emph{light}$ comes from these
young stars). 

The merging scenario provides another explanation for the age and metallicity trend with mass and concentration in the inner region. It has been shown analytically that positive correlations of mass with average age can be a natural consequence of hierarchical merging \citep{Nei06}. As well, detailed semi-analytic modeling of hierarchical merging that includes enhanced feedback processes such as AGN, yield more extended star formation histories in less massive galaxies,
leading to positive correlations of mass with average age and metallicity \citep{Col94,Bow06,Cro06,del06}. 

 Thus, the fact that we find 
mass-metallicity relations for the \emph{inner} regions of S0s does not 
discriminate between models where the bulge is formed through major 
mergers versus secular evolution models; they both predict age-mass and mass-metallicity relations for the bulges of S0s. Note that such relations also naturally result in a ``monolithic 
collapse'' scenario.

However, we see mass-metallicity relations in the outer regions as well. Can this point toward a particular formation mechanism?  \citet{Yoa08} propose that increased outflows in low mass disks at early times (in sub-units before assembling into a disk) allows a mass-metallicity relation in disks to be established early so that hierarchical merging may be able to explain the observed mass-metallicity relation in the outer regions of S0 galaxies. It is unclear whether such a relation can be explained by secular evolution considering star formation must be efficient enough in the outer regions for enrichment to occur. While the mass-metallicity relation observed in the outer regions of S0 galaxies may be a useful discriminant of formation mechanisms, predictions from both merger and secular evolution models have yet to consider this scenario.

The parallel between our finding of younger ages for low mass and low concentration S0s and a 
similar finding that late type spirals are younger than earlier types (M04) is consistent with a similar formation picture for S0s and spirals. 
We have demonstrated in \S\ref{sect:agetrend}  that the structural transition in ``transition'' galaxies correlates with a transition in SP properties between S0s and spirals; as with the spiral galaxies observed by M04, transition S0 galaxies do not show the tail of increasing ages that we observe for non-transition galaxies.  In fact, the majority of spiral galaxies are shown to have decreasing ages with radius (M04). The transition S0 galaxies thus appear to bridge the gap between S0 and spiral galaxies, supporting a formation connection between the two groups. Transition galaxies are likely candidates for descendants of spiral galaxies that have lost or exhausted their gas. However, the fact that many transition galaxies in our sample have large 
stellar masses is inconsistent with the proposed mass dichotomy in S0 
formation \citep{Jor94,Meh03,Barw07}, namely that only low mass S0s originated from 
spirals, while the high mass S0s have a formation history similar to ellipticals.

\subsection{Old Outskirts Galaxies}

We have found a subsample of galaxies with large positive age gradients and significantly old outer regions ($>$ 10 Gyr). These galaxies appear to form a separate evolutionary class from the rest of the sample, showing no preference for mass, concentration, or environment. Their physical appearance is largely featureless.  We have discussed potential observational and model dependent caveats for the result of old outskirts in \S\ref{sect:oops}. Assuming this result is robust, we explore tentative implications of galaxy evolution. 

Until we gather more information from light decompositions, we can only speculate on the nature of these old outskirts galaxies. The old populations in the outer regions either belong to a disk or a spheroid component. In the case of a spheroid membership, the galaxy may have a dominant spheroid population with an embedded disk so that the radial increase in ages is due to the change in luminosity-weighted ages from a disk-dominated to a spheroid-dominated region. \citet{Pel07}  found that the minor axis, and thus the spheroid component, of 24 early-type (Sa) galaxies has uniformly old SPs. It is this old spheroidal component that might be dominating in the outer regions of old outskirts S0s. In this scenario, old outskirts galaxies would be closely tied to elliptical galaxies. In fact, the latter are known to harbor small embedded disks \citep[][M10]{Kra08}, hence the old outskirts S0s would tend to straddle between ellipticals and S0 galaxies. Luminous red halos have also been observed around spiral galaxies \citep{Leq98,Zib04,Zak06}. However, \citet{Zak06} found that old populations cannot explain the red colors of many spiral halos without invoking an extremely bottom-heavy IMF. 

If, instead, the outer regions of old outskirts galaxies are dominated by a disk component, their stars would have formed long ago. Given uncertainties in SP modeling, setting an absolute age on the formation of the disk is difficult, but in old outskirts galaxies the stars dominating the light in the outer regions clearly formed long ago (possibly $>$ 10 Gyr). Moderately old disks ($\sim$ 7 Gyr)  have been predicted in theoretical models where the main infall phase precedes the onset of star formation \citep{Fer01}. As well,  \citet{Koo05} find several very red luminous disks at high redshift, implying that at $z=1$ not all massive disks are young; some old, massive S0s may have already existed in the field. For reference, stars formed at $z=1$ would be $\sim$ 7 Gyr old at the present epoch. The presence of a disk component implies that no major, disk-destroying event has occurred since the formation of the dynamically fragile stellar disk; stellar disks are thought to be destroyed in merger of galaxies with a mass ratio $>$ 1/4 \citep{MH96,Ste02}. Thus, the very old ages of the galaxies' outskirts (possibly $>$ 10 Gyr) place a constraint on models of hierarchical merging if the outskirts have a disky origin. 
  
Previous studies of spirals with low present day B/D suggest that hierarchical models may require no major merger since the formation of the stellar disk \citep{Wei09}. In principle, low B/D spirals may be produced through gas rich mergers, which produce a new disk in the merger remnant  \citep{Hop09,Rich09}. However, our observations of old outskirts galaxies cannot be explained through these mergers because new stars will necessarily form in the disk and would then dominate the light. Thus, the requirement of no major merger since $z \sim 2$ (using an age of 10 Gyr), if the outer regions are indeed a disk, is necessary to explain our observations.
 
 Alternately, old stars may have somehow been redistributed into the outer regions. However, because the integrated light is heavily weighted toward young SPs, the old light-weighted ages in the galaxies' outskirts imply that virtually no young stars exist in this region.  An accretion of old satellites to the outer disk may be possible. For example, \citet{deJ07} found that a stream around M83 is made of an old stellar population. Additionally, a migration of old stars from the center outward is predicted in the models of \citet{Ros08b} through spiral corotation resonance scattering, although the lack of spiral structure in S0 galaxies may make this difficult. 
   
 The large ages and low metallicities in the outer regions of old outskirts galaxies resemble those of thick disks observed around edge-on galaxies \citep{Dal02,Yoa08} and in the Milky Way \citep{Gil89}. Perhaps the populations in the old outskirts of galaxies  belong to a thick disk. Indeed  thick disks may form during a gas-rich merger event and an embedded thin disk forms afterward \citep{Rich09}. If a galaxy has its gas stripped shortly after this merger event, perhaps part or all of the young thin disk will be kept from forming and the galaxy will contain old outskirts.  Unfortunately, our galaxies are seen face-on so it is difficult to determine whether the outer populations belong to a thick disk.  We note that the \citet{Rich09} models invariably show a major starburst occurring during the merger. Thus, mergers like these must be old if they are to explain the outer populations of old outskirts galaxies.

Most disk/spiral galaxies show a decrease in light-weighted age with radius \citep{BDJ00, Mac04,Roe10}, a trend opposite to what is observed in old outskirts galaxies.   
The positive age gradients of old outskirts galaxies, which may be due to a difference in bulge and disk age and/or to an age gradient in the disk, can offer clues to their formation. 
If the bulge (or the population dominating the light in the bulge) formed after the disk, some mass in the bulge would need to be created without either destroying the fragile disk or adding star formation to it. Internal secular evolution in gaseous galaxies is expected to fuel star formation in a centralized disk. Internal secular evolution may play a role in creating the young bulges in old outskirts galaxies, but other processes would be required to explain the suppression of star formation in the disk.  Alternately, \citet{HM95} show that even a 10-to-1 merger can drive up to 50\% of the primary galaxy's gas into its center. Not only will this help to build up a young bulge, but it will suppress star formation in the disk. 

A positive age gradient in the disk component itself could be due to either the original formation of the disk or to a later event that transformed the galaxy into an S0.  Proposed theories of S0 formation through external gas removal processes \citep{Qui00,KM08,Kro08} call for an outside-in stripping of the galaxy. The resultant galaxy will have a lower light-weighted age inside the truncation radius than beyond it. Additionally, enhancement of star formation from several gas removal processes is expected in the central regions, again yielding an increase in age with radius. The events that are expected to bring about this age difference, however, are also expected to be short-lived with a timescale around 500 Myr \citep{Kro08}. To create the observed large positive age gradient, star formation must be sustained in the inner regions for a very long time, possibly through minor accretion. The age increase in old outskirts galaxies' disks may instead be due to the original formation of the disk. 

Positive age gradients in disks are not exclusive to S0 galaxies. \citet{Tay05} found that the outer regions of some late type spiral galaxies are redder and the inner regions are bluer than other galaxies in his sample, suggesting that this could be an indicator of outside-in formation, with significant recent star formation in the inner regions and minimal recent star formation in the outer regions. A positive age gradient is predicted in simulations of the dissipational collapse of gas embedded in a spherical dark matter halo through scattering of disk stars  \citep{Ros08}. Beyond a nominal radius, \citet{Ros08} find an increase in mean stellar age for their model disk galaxy.  Spiral arms at some point in the history of the host galaxy are a necessary component for this model.  

The scenario of the thick disk forming first may also be used to explain radial trends in old outskirts galaxies. Most likely, several processes are involved, i.e. perhaps the formation of the thick disk is halted through a partial stripping mechanism. Young disks embedded in the thick old disk and the outside-in disk growth in the models of \citet{Rich09} may explain the trend in radius toward older ages observed in old outskirts galaxies. 

In summary, the observed mass-metallicity trends in S0 galaxies are consistent with a merging scenario for the formation of the bulge, but the cause of the relation that we observe in the outer regions has yet to be considered in models. Our results, in particular those of transition galaxies that straddle S0 and spiral galaxies, are consistent with the formation of S0 galaxies through the transformation of spiral galaxy disks. It is possible that old outskirts galaxies require no major merger to have occurred in a very long time. These galaxies may harbor a thick disk in their outer region, while the thin disk was stripped early.  However, until the  structural components that make up these outer regions are resolved (e.g. disk versus spheroidal), we can only speculate about the nature of these galaxies.

\section{Conclusion}\label{sect:conclusion}

We have presented optical (SDSS $\emph{g}$ and $\emph{r}$) and NIR (J and/or H) surface photometry 
for a sample of 59 S0 galaxies covering a range in stellar mass and light
concentration.  Radial age and metallicity 
gradients out to at least 5 R$_{e}$ are derived from 
comparison of the observed $\emph{g}-$\emph{r} and $\emph{r}-H (and/or $\emph{r}-J) colors with SP models. 

We find an average central light-weighted age of $\sim$ 4 Gyr and central 
metallicity of [Z/H] $\sim$ 0.2 dex. For most of the galaxies in our sample, a large negative metallicity gradient is found with radius, with an average of 
$\Delta$[Z/H]$/$$\Delta$log(r/R$_e$) $=$ -0.6 $\pm$ 0.5. Radial age trends are more 
heterogeneous. An increase 
in age with radius is found for 58\% of our sample, a decrease for 19\%, and 
little change for 23\% of it. 

The outer regions of 14 out of 59 S0 galaxies have very old light-weighted ages ($>$ 10 Gyr) and also exhibit large increases in light-weighted age from the center outward. 
These old outskirts galaxies are found in a range of environments, masses, 
and concentrations. None of these galaxies
shows nascent spiral structure. Determining the 
structural component (i.e. disk versus spheroidal) that makes up the outer 
region is a necessary step to further probe the formation history of these 
galaxies.  Assuming that the outer region is a disk, old outskirts galaxies may be showing the thick disk in the outer region, while the thin disk was stripped early. It is possible that old outskirts galaxies require no major merger to have occurred in a very long time. 

We find that mean light-weighted age and metallicity correlate with both mass and concentration; 
galaxies with both lower 
mass and lower concentration have, on average, younger central ages and lower metallicities than other systems  
in our sample. The mass-metallicity relation extends  to the outer regions of the galaxy. The observed mass-metallicity trends in S0 galaxies are consistent with a merging formation of these galaxies. The combination of a statistically incomplete sample regarding environment (e.g., all of the observed lowest mass galaxies are in the Virgo cluster high density environment) and a scarcity of high density environment galaxies makes it difficult to draw conclusions regarding the effect of environment on age or metallicity.

We wish to thank  S. Charlot and G. Bruzual for providing their new models, J. Miner for providing density calculations, J. Roediger and S. Kannappan for stimulating discussions, and Brent Tully who made all of our NIR observations at the UH 2.2m possible. We thank the referee for numerous thoughtful comments that led to a much improved presentation. This study was partially funded by NSF grant AST 04-06443 to the University of North Carolina. L.C. acknowledges the support of the Linda Dykstra Science Dissertation Fellowship and 
S.C. acknowledges the support of NSERC through Discovery grant. This research has made use of the NASA/IPAC Extragalactic Database (NED) which is operated by the Jet Propulsion Laboratory, California Institute of Technology, under contract with the National Aeronautics and Space Administration.

\bibliographystyle{mn2e.bst}	
\bibliography{biblio} 

\clearpage

\input{stub.tab1.tex}

\input{stub.tabobserve}

\clearpage

\input{stub.tab2.tex}

\input{stub.tab3.tex}

\clearpage

\begin{table}
\centering
\caption{Binning Scheme}
\begin{tabular}{ll}
\hline

Bin&Radial range\\
\hline
1 & $0.0\less$r/R$_{e}$$\less0.5$\\
2 & $0.5\less$r/R$_{e}$$\less1.5$\\
3 & $1.5\less$r/R$_{e}$$\less2.5$\\
4 & $2.5\less$r/R$_{e}$$\less3.5$\\
5 & $3.5\less$r/R$_{e}$$\less4.5$\\
6 & $4.5\less$r/R$_{e}$$\less5.5$\\
\hline
\end{tabular}
\end{table}

\input{stub.tab4.tex}

\input{stattab.tex}

\clearpage

\clearpage

\clearpage

\clearpage

\clearpage

\clearpage

\clearpage

\clearpage

\end{document}

%% file: stub.tab1.tex
\pagestyle{empty}

\begin{table*}
\begin{minipage}{140mm}

\caption{Galaxy Characteristics}

\begin{tabular}{llllllllrll}
\hline

Galaxy&Morph$^{\emph{i}}$
&a$^{\emph{j}}$
&b$^{\emph{j}}$
&V$_r$$^{\emph{j}}$
&Distance$^{\emph{k}}$
&Incl.$^{\emph{i}}$
&Type$^{\emph{l}}$
\\
\multicolumn{1}{l}{}&\multicolumn{1}{l}{}&
\multicolumn{1}{l}{(')}&\multicolumn{1}{l}{(')}&
\multicolumn{1}{l}{(\kms)}&\multicolumn{1}{l}{(Mpc)}&
\multicolumn{1}{l}{($^\circ$)}&\multicolumn{1}{l}{}
\\
\hline

UGC04330& SB0& 1.1& 1.0& 4865& 69.4& 34.6& B\\
UGC04596& S0& 1.2& 1.0& 9439& 131.7& 25.4& T\\
UGC04599& S0& 2.1& 2.1& 2072& 31.8& 18.3& T\\
UGC04631& S0& 1.1& 1.0& 4159& 60.6& 24.3& N\\
UGC04639& S0?& 1.4& 1.3& 8556& 120.0& 23.3& N\\
UGC04737& S0?& 0.8& 0.5& 3813& 56.0& 53.3& N\\
UGC04869& S0?& 2.0& 0.7& 6889& 97.8& 62.9& T,B\\
UGC04901& S0-A?& 1.1& 1.1& 8424& 118.7& 28.9& N\\
UGC04910& S0& 1.0& 0.6& 8353& 117.8& 28.9& N\\
UGC04916& S0& 1.2& 0.9& 8785& 123.6& 36.2& N\\
\end{tabular}

\medskip
The full table appears in the electronic edition of the Journal.\\
 \\
$^{\emph{i}}$ From the Uppsala General Catalog (UGC) unless otherwise noted\\
$^{\emph{j}}$ Major and minor diameter, and radial velocity, from the NASA/IPEC Extragalactic Database (NED)\\
$^{\emph{k}} $ Distances are corrected for Virgo flow and the Great Attractor, from NED\\
$^{\emph{l}}$ T (transition galaxy), B (barred, from Hyperleda), N (normal)\\
$^{\emph{m}}$ From NED if morphology; from Hyperleda if inclination\\

\end{minipage}
\end{table*}

%% file: stub.tabobserve.tex
\pagestyle{empty}

\begin{table*}
\begin{minipage}{140mm}

\caption{UH 2.2-m Observations}

\begin{tabular}{lllllr}
\hline

Galaxy&T(sec)\footnote{Total exposure time in seconds}
&date&Z.P.(mag)\footnote{Zero-point magnitude}
&$\sigma$ \footnote{Standard deviation in the derived zero-points for that field}
&NS\footnote{Number of standard stars in the field}
\\
\hline

UGC04330J&480&  Apr 2007&-1.916& 0.037 & 10\\
UGC04596H&480&  Mar 2008&-1.731 & 0.004 & 3\\
UGC04599J&480&  Apr 2007&-1.713& 0.034 & 11\\
UGC04631J&480&  Apr 2007&-1.707& 0.037 & 7\\
UGC04639J&480&  Apr 2007&-1.628& 0.030 & 9\\
UGC04737H&480&  Mar 2008&-1.819& 0.201 & 10\\
UGC04869H&480&  Mar 2008&-1.853& 0.056 & 0\\
UGC04901J&480&  Apr 2007&-1.596& 0.025 & 9\\
UGC04910J&480&  Apr 2007&-1.579& 0.057 & 7\\
UGC04916J&480&  Apr 2007&-1.694& 0.018 & 7\\

\end{tabular}

\medskip
The full table appears in the electronic edition of the Journal.

\end{minipage}
\end{table*}

%% file: stub.tab2.tex
\pagestyle{empty}

\begin{table*}
\begin{minipage}{200mm}

\caption{Derived Radial Quantities}
\begin{tabular}{llllllllllllll}
\hline

\multicolumn{1}{l}{}&\multicolumn{1}{l}{}&
\multicolumn{4}{c}{Inner}&
\multicolumn{1}{l}{}&
\multicolumn{4}{c}{Outer}\\
Name&cutoff \footnote{Outer radial cutoff in arcseconds} (")
&age(Gyr)&+$\sigma$&-$\sigma$
&[Z/H]&+$\sigma$&-$\sigma$
&age(Gyr)&+$\sigma$&-$\sigma$
&[Z/H]&+$\sigma$&-$\sigma$
\\
\hline

\multicolumn{1}{l}{H band derived }\\
UGC04596 & 38 & 10.53 &  0.81 &  0.32 &  0.06 &  0.01 &  0.00 &  2.49 &  0.15 &  0.14 & -0.47 &  0.03 &  0.04 \\
UGC04737 & 62 &  4.69 &  0.20 &  0.18 &  0.22 &  0.01 &  0.01 &  7.31 &  1.57 &  1.13 & -0.22 &  0.05 &  0.06 \\
UGC04869 & 100 &  3.97 &   -- &    -- &  0.50 &    -- &    -- &  2.88 &  0.72 &  0.25 &  0.03 &  0.05 &  0.05 \\ 
UGC05094 & 60 &  3.48 &  0.15 &    -- &  0.48 &  0.02 &    -- &  5.51 &  3.64 &  2.19 &  0.04 &  0.07 &  0.06 \\ 
UGC05182 & 50 &  5.83 &  0.43 &  0.21 &  0.23 &  0.01 &  0.01 & 13.19 &    -- &  2.35 & -0.50 &    -- &  0.05 \\ 
\multicolumn{1}{l}{J band derived}\\
UGC04330 & 43 &  2.08 &    -- &    -- &  0.50 &    -- &    -- &  3.00 &  0.77 &  0.21 &  0.07 &  0.05 &  0.05 \\ 
UGC04599 & 58 &  2.73 &  0.05 &  0.06 & -0.01 &  0.02 &  0.02 &  1.05 &  0.02 &  0.02 & -0.48 &  0.02 &  0.02 \\ 
UGC04631 & 36 &  3.90 &  0.42 &  0.16 &  0.19 &  0.02 &  0.01 & 13.69 &    -- &  1.08 & -0.69 &    -- &  0.23 \\ 
UGC04639 & 58 &  4.76 &  0.27 &  0.24 &  0.28 &  0.02 &  0.02 &  2.17 &  0.33 &  0.25 &  0.19 &  0.03 &  0.04 \\ 
UGC04901 & 90 &  3.49 &  0.08 &  0.03 &  0.40 &  0.01 &  0.01 &  2.50 &    -- &    -- &  0.47 &    -- &    -- \\

\end{tabular}

\medskip
The full table appears in the electronic edition of the Journal.

\end{minipage}
\end{table*}

%% file: stub.tab3.tex
\pagestyle{empty}

\begin{table*}
\begin{minipage}{140mm}
\caption{Derived Total Quantities}
\begin{tabular}{lllllllllll}

\hline

Name&r$_{tot}$&g$_{tot}$&H$_{tot}$&J$_{tot}$&Re$_r$(")&g-r$_{tot}$&r-H$_{tot}$
&r-J$_{tot}$
\\

 UGC04596  & 14.0 & 14.7 & 11.4 &  -- &  9.9 &  0.7 &  2.6 & --  \\ 
 UGC04737  & 13.2 & 14.0 & 10.4 &  -- &  8.2 &  0.8 &  2.8 & --  \\ 
 UGC04869  & 12.9 & 13.8 & 10.0 &  -- & 11.9 &  0.8 &  3.0 & --  \\ 
 UGC05094  & 14.0 & 14.8 & 11.2 &  -- &  8.3 &  0.8 &  2.8 & --  \\ 
 UGC05182  & 13.0 & 13.8 & 10.4 &  -- & 10.6 &  0.9 &  2.6 & --  \\ 
 UGC05403  & 13.3 & 14.1 & 10.6 &  -- & 10.7 &  0.8 &  2.7 & --  \\ 
 UGC05419  & 12.3 & 13.0 &  9.7 &  -- & 19.0 &  0.7 &  2.6 & --  \\ 
 UGC05568  & 12.1 & 12.8 &  9.7 &  -- & 15.3 &  0.7 &  2.4 & --  \\ 
 UGC05766  & 12.2 & 12.9 &  9.6 &  -- & 17.2 &  0.8 &  2.6 & --  \\ 
 UGC08886  & 12.9 & 13.7 & 10.1 &  -- &  6.1 &  0.8 &  2.8 & --  \\ 

 \end{tabular}

\medskip
The full table appears in the electronic edition of the Journal.

\end{minipage}
\end{table*}

%% file: stub.tab4.tex
\pagestyle{empty}

\begin{table*}
\begin{minipage}{140mm}
\caption{Derived Global Properties}

\begin{tabular}{lclr}

\hline

Name&Concentration&Mass&Local Density 
\\

 & &(M$_{\sun}$)&(log \mpc)
\\

\hline

UGC04330 & 4.4 & 1.22e+11 & 0.40  \\
UGC04596 & 5.3 & 1.02e+11 & -1.41  \\
UGC04599 & 4.4 & 6.97e+09 & 0.29  \\
UGC04631 & 4.9 & 3.88e+10 & 0.45  \\
UGC04639 & 4.7 & 2.29e+11 & -1.22  \\
UGC04737 & 4.7 & 4.92e+10 & -0.07  \\
UGC04869 & 4.7 & 1.87e+11 & -0.29  \\
UGC04901 & 4.2 & 3.62e+11 & 0.43  \\
UGC04910 & 4.9 & 3.24e+11 & 0.94  \\
UGC04916 & 5.3 & 1.95e+11 & -0.72  \\
\end{tabular}

\medskip
The full table appears in the electronic edition of the Journal.

\end{minipage}
\end{table*}

%% file: stattab.tex
\pagestyle{empty}

\begin{table*}
\begin{minipage}{140mm}

\caption{Results from Statistical Analysis}

\begin{tabular}{lll}
\hline
Groups  &Result \\
\hline
\multicolumn{2}{l}{ Age Trends}\\
inner region ages from J-band and H-band for galaxies in common & 0.81\\
outer region ages from J-band and H-band for galaxies in common & 0.81\\
inner ages of high and low concentration galaxies&0.03\\
outer ages of high and low concentration galaxies & 0.20\\
inner ages of high and low mass galaxies  & 0.03\\
outer ages of high and low mass galaxies  &0.06\\
ages for all bins of high and low concentration galaxies	 & 1.90E-4\\
ages for all bins of high and low mass galaxies	&  1.70E-5\\
ages for all bins of high and low mass galaxies with low concentration&0.09\\
ages for all bins of high and low mass galaxies with high concentration & 0.02\\
ages for all bins of high and low concentration galaxies with low mass & 0.19\\
ages for all bins of high and low concentration galaxies with high mass  & 0.04\\
age differences of transition and featureless disk galaxies & 1.20E-4\\
outer ages of high and low concentration non-old-outskirts galaxies& 0.34\\
outer ages of high and low mass non-old-outskirts galaxies  &0.05\\
outer ages of high and low density non-old-outskirts galaxies  & 0.99\\
mass of old-outskirts and non-old-outskirts galaxies & 0.71\\
concentration of old-outskirts and non-old-outskirts galaxies & 0.91\\
environmental density of old-outskirts and non-old-outskirts galaxies & 0.36\\
\multicolumn{2}{l}{ Metallicity Trends}\\
inner region metallicity from J-band and H-band for galaxies in common & 0.81\\
outer region metallicity from J-band and H-band for galaxies in common & 0.81\\
inner metallicity of high and low concentration galaxies & 7.49E-4\\
outer metallicity of high and low concentration galaxies & 0.10\\
inner metallicity of high and low mass galaxies & 6.77E-5\\
outer metallicity of high and low mass galaxies& 0.01\\
metallicity for all bins of high and low concentration galaxies	 & 0.09\\
metallicity for all bins of high and low mass galaxies	&  8.10E-4\\
metallicity for all bins of high and low mass galaxies with low concentration&5.90E-6\\
metallicity for all bins of high and low mass galaxies with high concentration  & 0.53\\
metallicity for all bins of high and low concentration galaxies with low mass & 9.1E-4\\
metallicity for all bins of high and low concentration galaxies with high mass & 0.06\\
inner metallicity of high and low mass galaxies with low concentration &2.5E-4\\
inner metallicity of high and low mass galaxies with high concentration & 0.20\\
inner metallicity of high and low concentration galaxies with low mass  & 2.5E-4\\
inner metallicity of high and low concentration galaxies with high mass& 0.68\\
metallicity gradient ($\Delta [M/H]/\Delta log(r)$) of high and low mass  galaxies  & 0.30\\
metallicity gradient ($\Delta [M/H]/\Delta log(r)$) of high and low concentration galaxies  &  0.30\\
\hline
\end{tabular}

\medskip
Probability from Kolmogorov-Smirnov test that the two groups are drawn from the same population.
Separation of concentration groups is at $C_{28}$ $=$ 4.7, separation of mass is at 1$\times10^{11} $ M$_{\sun}$, and separation of density is at -0.3.

\end{minipage}
\end{table*}

%% file: s0_mn.bbl
\begin{thebibliography}{}

\bibitem[\protect\citeauthoryear{{Balcells} \& {Peletier}}{{Balcells} \&
  {Peletier}}{1994}]{BP94}
{Balcells} M.,  {Peletier} R.~F.,  1994, \aj, 107, 135

\bibitem[\protect\citeauthoryear{{Barnes}}{{Barnes}}{2002}]{Bar02}
{Barnes} J.~E.,  2002, \mnras, 333, 481

\bibitem[\protect\citeauthoryear{{Barway}, {Kembhavi}, {Wadadekar}, {Ravikumar}
  \& {Mayya}}{{Barway} et~al.}{2007}]{Barw07}
{Barway} S.,  {Kembhavi} A.,  {Wadadekar} Y.,  {Ravikumar} C.~D.,    {Mayya}
  Y.~D.,  2007, \apjl, 661, L37

\bibitem[\protect\citeauthoryear{{Bedregal}, {Arag{\'o}n-Salamanca},
  {Merrifield} \& {Cardiel}}{{Bedregal} et~al.}{2008}]{Bed08}
{Bedregal} A.~G.,  {Arag{\'o}n-Salamanca} A.,  {Merrifield} M.~R.,    {Cardiel}
  N.,  2008, \mnras, 387, 660

\bibitem[\protect\citeauthoryear{{Bell} \& {de Jong}}{{Bell} \& {de
  Jong}}{2000}]{BDJ00}
{Bell} E.~F.,  {de Jong} R.~S.,  2000, \mnras, 312, 497

\bibitem[\protect\citeauthoryear{{Bell}, {McIntosh}, {Katz} \&
  {Weinberg}}{{Bell} et~al.}{2003}]{Bel03}
{Bell} E.~F.,  {McIntosh} D.~H.,  {Katz} N.,    {Weinberg} M.~D.,  2003, \apjs,
  149, 289

\bibitem[\protect\citeauthoryear{{Bell}, {Wolf}, {Meisenheimer}, {Rix},
  {Borch}, {Dye}, {Kleinheinrich} \& {Plus 2 more}}{{Bell}
  et~al.}{2004}]{Bel04}
{Bell} E.~F.,  {Wolf} C.,  {Meisenheimer} K.,  {Rix} H.-W.,  {Borch} A.,  {Dye}
  S.,  {Kleinheinrich} M.,    {Plus 2 more} 2004, \apj, 608, 752

\bibitem[\protect\citeauthoryear{{Bertelli}, {Girardi}, {Marigo} \&
  {Nasi}}{{Bertelli} et~al.}{2008}]{Ber08}
{Bertelli} G.,  {Girardi} L.,  {Marigo} P.,    {Nasi} E.,  2008, \aap, 484, 815

\bibitem[\protect\citeauthoryear{{Boselli}, {Boissier}, {Cortese} \&
  {Gavazzi}}{{Boselli} et~al.}{2008}]{Bos08}
{Boselli} A.,  {Boissier} S.,  {Cortese} L.,    {Gavazzi} G.,  2008, \apj, 674,
  742

\bibitem[\protect\citeauthoryear{{Bothun} \& {Gregg}}{{Bothun} \&
  {Gregg}}{1990}]{BG90}
{Bothun} G.~D.,  {Gregg} M.~D.,  1990, \apj, 350, 73

\bibitem[\protect\citeauthoryear{{Bower}, {Benson}, {Malbon}, {Helly}, {Frenk},
  {Baugh}, {Cole} \& {Lacey}}{{Bower} et~al.}{2006}]{Bow06}
{Bower} R.~G.,  {Benson} A.~J.,  {Malbon} R.,  {Helly} J.~C.,  {Frenk} C.~S.,
  {Baugh} C.~M.,  {Cole} S.,    {Lacey} C.~G.,  2006, \mnras, 370, 645

\bibitem[\protect\citeauthoryear{{Bruzual} \& {Charlot}}{{Bruzual} \&
  {Charlot}}{2003}]{BC03}
{Bruzual} G.,  {Charlot} S.,  2003, \mnras, 344, 1000

\bibitem[\protect\citeauthoryear{{Butcher} \& {Oemler} Jr.}{{Butcher} \&
  {Oemler}}{1978}]{BO78}
{Butcher} H.,  {Oemler} Jr. A.,  1978, \apj, 226, 559

\bibitem[\protect\citeauthoryear{{Byrd} \& {Valtonen}}{{Byrd} \&
  {Valtonen}}{1990}]{BV90}
{Byrd} G.,  {Valtonen} M.,  1990, \apj, 350, 89

\bibitem[\protect\citeauthoryear{{Caldwell}}{{Caldwell}}{1983}]{Cal83}
{Caldwell} N.,  1983, \apj, 268, 90

\bibitem[\protect\citeauthoryear{{Caldwell}, {Rose} \& {Concannon}}{{Caldwell}
  et~al.}{2003}]{CRC03}
{Caldwell} N.,  {Rose} J.~A.,    {Concannon} K.~D.,  2003, \aj, 125, 2891

\bibitem[\protect\citeauthoryear{{Charlot} \& {Bruzual}}{{Charlot} \&
  {Bruzual}}{2010}]{CB09}
{Charlot} S.,  {Bruzual} G.,  2010, in preparation

\bibitem[\protect\citeauthoryear{{Cole}, {Aragon-Salamanca}, {Frenk}, {Navarro}
  \& {Zepf}}{{Cole} et~al.}{1994}]{Col94}
{Cole} S.,  {Aragon-Salamanca} A.,  {Frenk} C.~S.,  {Navarro} J.~F.,    {Zepf}
  S.~E.,  1994, \mnras, 271, 781

\bibitem[\protect\citeauthoryear{{Combes}}{{Combes}}{2000}]{Com00}
{Combes} F.,  2000, in {Combes} F.,  {Mamon} G.~A.,   {Charmandaris} V.,  eds,
  Dynamics of Galaxies: from the Early Universe to the Present Vol.~197 of
  Astronomical Society of the Pacific Conference Series, {Bar-driven Galaxy
  Evolution and Time-scales to Feed AGN}.
pp 15--+

\bibitem[\protect\citeauthoryear{{Conroy}, {White} \& {Gunn}}{{Conroy}
  et~al.}{2009}]{Con09}
{Conroy} C.,  {White} M.,    {Gunn} J.~E.,  2009, ArXiv e-prints

\bibitem[\protect\citeauthoryear{{Cooper}, {Tremonti}, {Newman} \&
  {Zabludoff}}{{Cooper} et~al.}{2008}]{Coo08}
{Cooper} M.~C.,  {Tremonti} C.~A.,  {Newman} J.~A.,    {Zabludoff} A.~I.,
  2008, \mnras, 390, 245

\bibitem[\protect\citeauthoryear{{Courteau}}{{Courteau}}{1996}]{Cou96}
{Courteau} S.,  1996, \apjs, 103, 363

\bibitem[\protect\citeauthoryear{{Courteau}, {de Jong} \& {Broeils}}{{Courteau}
  et~al.}{1996}]{Cou96b}
{Courteau} S.,  {de Jong} R.~S.,    {Broeils} A.~H.,  1996, \apjl, 457, L73+

\bibitem[\protect\citeauthoryear{{Croton}, {Springel}, {White}, {De Lucia},
  {Frenk}, {Gao}, {Jenkins} \& {plus 3 more}}{{Croton} et~al.}{2006}]{Cro06}
{Croton} D.~J.,  {Springel} V.,  {White} S.~D.~M.,  {De Lucia} G.,  {Frenk}
  C.~S.,  {Gao} L.,  {Jenkins} A.,    {plus 3 more} 2006, \mnras, 365, 11

\bibitem[\protect\citeauthoryear{{Dalcanton} \& {Bernstein}}{{Dalcanton} \&
  {Bernstein}}{2002}]{Dal02}
{Dalcanton} J.~J.,  {Bernstein} R.~A.,  2002, \aj, 124, 1328

\bibitem[\protect\citeauthoryear{{de Jong} \& {Davies}}{{de Jong} \&
  {Davies}}{1997}]{DJD97}
{de Jong} R.~S.,  {Davies} R.~L.,  1997, \mnras, 285, L1

\bibitem[\protect\citeauthoryear{{de Jong}, {Seth}, {Bell}, {Brown}, {Bullock},
  {Courteau}, {Dalcanton} \& {plus 6 more}}{{de Jong} et~al.}{2007}]{deJ07}
{de Jong} R.~S.,  {Seth} A.~C.,  {Bell} E.~F.,  {Brown} T.~M.,  {Bullock}
  J.~S.,  {Courteau} S.,  {Dalcanton} J.~J.,    {plus 6 more} 2007, in
  {A.~Vazdekis \& R.~F.~Peletier} ed., IAU Symposium Vol.~241 of IAU Symposium,
  {GHOSTS: The Resolved Stellar Outskirts of Massive Disk Galaxies}.
pp 503--504

\bibitem[\protect\citeauthoryear{{De Lucia}, {Springel}, {White}, {Croton} \&
  {Kauffmann}}{{De Lucia} et~al.}{2006}]{del06}
{De Lucia} G.,  {Springel} V.,  {White} S.~D.~M.,  {Croton} D.,    {Kauffmann}
  G.,  2006, \mnras, 366, 499

\bibitem[\protect\citeauthoryear{{Dressler}}{{Dressler}}{1980}]{Dre80}
{Dressler} A.,  1980, \apj, 236, 351

\bibitem[\protect\citeauthoryear{{Dressler} \& {Gunn}}{{Dressler} \&
  {Gunn}}{1983}]{DG83}
{Dressler} A.,  {Gunn} J.~E.,  1983, \apj, 270, 7

\bibitem[\protect\citeauthoryear{{Ellis}, {Abraham} \& {Dickinson}}{{Ellis}
  et~al.}{2001}]{Ell01}
{Ellis} R.~S.,  {Abraham} R.~G.,    {Dickinson} M.,  2001, \apj, 551, 111

\bibitem[\protect\citeauthoryear{{Eminian}, {Kauffmann}, {Charlot}, {Wild},
  {Bruzual}, {Rettura} \& {Loveday}}{{Eminian} et~al.}{2008}]{Emi07}
{Eminian} C.,  {Kauffmann} G.,  {Charlot} S.,  {Wild} V.,  {Bruzual} G.,
  {Rettura} A.,    {Loveday} J.,  2008, \mnras, 384, 930

\bibitem[\protect\citeauthoryear{{Falco}, {Kurtz}, {Geller}, {Huchra},
  {Peters}, {Berlind}, {Mink} \& {Plus 2 more}}{{Falco} et~al.}{1999}]{Fal99}
{Falco} E.~E.,  {Kurtz} M.~J.,  {Geller} M.~J.,  {Huchra} J.~P.,  {Peters} J.,
  {Berlind} P.,  {Mink} D.~J.,    {Plus 2 more} 1999, \pasp, 111, 438

\bibitem[\protect\citeauthoryear{{Ferguson} \& {Clarke}}{{Ferguson} \&
  {Clarke}}{2001}]{Fer01}
{Ferguson} A.~M.~N.,  {Clarke} C.~J.,  2001, \mnras, 325, 781

\bibitem[\protect\citeauthoryear{{Fisher}, {Franx} \& {Illingworth}}{{Fisher}
  et~al.}{1996}]{Fis96}
{Fisher} D.,  {Franx} M.,    {Illingworth} G.,  1996, \apj, 459, 110

\bibitem[\protect\citeauthoryear{{Freeman}}{{Freeman}}{1970}]{Fre70}
{Freeman} K.~C.,  1970, \apj, 160, 811

\bibitem[\protect\citeauthoryear{{Fritze v.~Alvensleben}}{{Fritze
  v.~Alvensleben}}{2004}]{Fri04}
{Fritze v.~Alvensleben} U.,  2004, in {D.~L.~Block, I.~Puerari, K.~C.~Freeman,
  R.~Groess, \& E.~K.~Block } ed., Penetrating Bars Through Masks of Cosmic
  Dust Vol.~319 of Astrophysics and Space Science Library, {On the Origin of SO
  Galaxies}.
pp 81--+

\bibitem[\protect\citeauthoryear{{Garnett}}{{Garnett}}{2002}]{Gar02}
{Garnett} D.~R.,  2002, \apj, 581, 1019

\bibitem[\protect\citeauthoryear{{Gilmore}, {Wyse} \& {Kuijken}}{{Gilmore}
  et~al.}{1989}]{Gil89}
{Gilmore} G.,  {Wyse} R.~F.~G.,    {Kuijken} K.,  1989, \araa, 27, 555

\bibitem[\protect\citeauthoryear{{Graves}, {Faber} \& {Schiavon}}{{Graves}
  et~al.}{2009}]{Gra09}
{Graves} G.~J.,  {Faber} S.~M.,    {Schiavon} R.~P.,  2009, \apj, 693, 486

\bibitem[\protect\citeauthoryear{{Hernquist} \& {Mihos}}{{Hernquist} \&
  {Mihos}}{1995}]{HM95}
{Hernquist} L.,  {Mihos} J.~C.,  1995, \apj, 448, 41

\bibitem[\protect\citeauthoryear{{Hinshaw}, {Weiland}, {Hill}, {Odegard},
  {Larson}, {Bennett}, {Dunkley} \& {plus 14 more}}{{Hinshaw}
  et~al.}{2009}]{Hin09}
{Hinshaw} G.,  {Weiland} J.~L.,  {Hill} R.~S.,  {Odegard} N.,  {Larson} D.,
  {Bennett} C.~L.,  {Dunkley} J.,    {plus 14 more} 2009, \apjs, 180, 225

\bibitem[\protect\citeauthoryear{{Hopkins}, {Somerville}, {Cox}, {Hernquist},
  {Jogee}, {Kere{\v s}}, {Ma}, {Robertson} \& {Stewart}}{{Hopkins}
  et~al.}{2009}]{Hop09}
{Hopkins} P.~F.,  {Somerville} R.~S.,  {Cox} T.~J.,  {Hernquist} L.,  {Jogee}
  S.,  {Kere{\v s}} D.,  {Ma} C.-P.,  {Robertson} B.,    {Stewart} K.,  2009,
  \mnras, 397, 802

\bibitem[\protect\citeauthoryear{{Icke}}{{Icke}}{1985}]{Ick85}
{Icke} V.,  1985, \aap, 144, 115

\bibitem[\protect\citeauthoryear{{Jorgensen} \& {Franx}}{{Jorgensen} \&
  {Franx}}{1994}]{Jor94}
{Jorgensen} I.,  {Franx} M.,  1994, \apj, 433, 553

\bibitem[\protect\citeauthoryear{{Kannappan}, {Guie} \& {Baker}}{{Kannappan}
  et~al.}{2009a}]{Kan08}
{Kannappan} S.,  {Guie} J.,    {Baker} A.,  2009a, \aj, submitted

\bibitem[\protect\citeauthoryear{{Kannappan}, {Guie} \& {Baker}}{{Kannappan}
  et~al.}{2009b}]{Kan09}
{Kannappan} S.~J.,  {Guie} J.~M.,    {Baker} A.~J.,  2009b, \aj, 138, 579

\bibitem[\protect\citeauthoryear{{Kannappan}, {Jansen} \& {Barton}}{{Kannappan}
  et~al.}{2004}]{Kan04}
{Kannappan} S.~J.,  {Jansen} R.~A.,    {Barton} E.~J.,  2004, \aj, 127, 1371

\bibitem[\protect\citeauthoryear{{Kauffmann}, {Heckman}, {White}, {Charlot},
  {Tremonti}, {Brinchmann}, {Bruzual} \& {Plus 15 more}}{{Kauffmann}
  et~al.}{2003}]{Kau03}
{Kauffmann} G.,  {Heckman} T.~M.,  {White} S.~D.~M.,  {Charlot} S.,  {Tremonti}
  C.,  {Brinchmann} J.,  {Bruzual} G.,    {Plus 15 more} 2003, \mnras, 341, 33

\bibitem[\protect\citeauthoryear{{Kawata} \& {Mulchaey}}{{Kawata} \&
  {Mulchaey}}{2008}]{KM08}
{Kawata} D.,  {Mulchaey} J.~S.,  2008, \apjl, 672, L103

\bibitem[\protect\citeauthoryear{{Kent}}{{Kent}}{1985}]{Ken85}
{Kent} S.~M.,  1985, \apjs, 59, 115

\bibitem[\protect\citeauthoryear{{Koo}, {Simard}, {Willmer}, {Gebhardt},
  {Bouwens}, {Kauffmann}, {Crosby} \& {plus 8 more}}{{Koo}
  et~al.}{2005}]{Koo05}
{Koo} D.~C.,  {Simard} L.,  {Willmer} C.~N.~A.,  {Gebhardt} K.,  {Bouwens}
  R.~J.,  {Kauffmann} G.,  {Crosby} T.,    {plus 8 more} 2005, \apjs, 157, 175

\bibitem[\protect\citeauthoryear{{Kormendy} \& {Kennicutt} Jr.}{{Kormendy} \&
  {Kennicutt}}{2004}]{Kor04}
{Kormendy} J.,  {Kennicutt} Jr. R.~C.,  2004, \araa, 42, 603

\bibitem[\protect\citeauthoryear{{Krajnovi{\'c}}, {Bacon}, {Cappellari},
  {Davies}, {de Zeeuw}, {Emsellem}, {Falc{\'o}n-Barroso} \& {plus 6
  more}}{{Krajnovi{\'c}} et~al.}{2008}]{Kra08}
{Krajnovi{\'c}} D.,  {Bacon} R.,  {Cappellari} M.,  {Davies} R.~L.,  {de Zeeuw}
  P.~T.,  {Emsellem} E.,  {Falc{\'o}n-Barroso} J.,    {plus 6 more} 2008,
  \mnras, 390, 93

\bibitem[\protect\citeauthoryear{{Kronberger}, {Kapferer}, {Ferrari},
  {Unterguggenberger} \& {Schindler}}{{Kronberger} et~al.}{2008}]{Kro08}
{Kronberger} T.,  {Kapferer} W.,  {Ferrari} C.,  {Unterguggenberger} S.,
  {Schindler} S.,  2008, \aap, 481, 337

\bibitem[\protect\citeauthoryear{{Kuntschner}, {Emsellem}, {Bacon}, {Bureau},
  {Cappellari}, {Davies}, {de Zeeuw} \& {plus 5 more}}{{Kuntschner}
  et~al.}{2006}]{Kun06}
{Kuntschner} H.,  {Emsellem} E.,  {Bacon} R.,  {Bureau} M.,  {Cappellari} M.,
  {Davies} R.~L.,  {de Zeeuw} P.~T.,    {plus 5 more} 2006, \mnras, 369, 497

\bibitem[\protect\citeauthoryear{{Le Borgne}, {Bruzual}, {Pell{\'o}}, {Lan{\c
  c}on}, {Rocca-Volmerange}, {Sanahuja}, {Schaerer}, {Soubiran} \&
  {V{\'{\i}}lchez-G{\'o}mez}}{{Le Borgne} et~al.}{2003}]{LeB03}
{Le Borgne} J.,  {Bruzual} G.,  {Pell{\'o}} R.,  {Lan{\c c}on} A.,
  {Rocca-Volmerange} B.,  {Sanahuja} B.,  {Schaerer} D.,  {Soubiran} C.,
  {V{\'{\i}}lchez-G{\'o}mez} R.,  2003, \aap, 402, 433

\bibitem[\protect\citeauthoryear{{Lee}, {Worthey}, {Dotter}, {Chaboyer},
  {Jevremovi{\'c}}, {Baron}, {Briley}, {Ferguson}, {Coelho} \& {Trager}}{{Lee}
  et~al.}{2009}]{Lee09}
{Lee} H.,  {Worthey} G.,  {Dotter} A.,  {Chaboyer} B.,  {Jevremovi{\'c}} D.,
  {Baron} E.,  {Briley} M.~M.,  {Ferguson} J.~W.,  {Coelho} P.,    {Trager}
  S.~C.,  2009, \apj, 694, 902

\bibitem[\protect\citeauthoryear{{Lequeux}, {Combes}, {Dantel-Fort},
  {Cuillandre}, {Fort} \& {Mellier}}{{Lequeux} et~al.}{1998}]{Leq98}
{Lequeux} J.,  {Combes} F.,  {Dantel-Fort} M.,  {Cuillandre} J.,  {Fort} B.,
  {Mellier} Y.,  1998, \aap, 334, L9

\bibitem[\protect\citeauthoryear{{MacArthur}, {McDonald}, {Courteau} \&
  {Gonzalez}}{{MacArthur} et~al.}{2010}]{Mac10}
{MacArthur} L.,  {McDonald} M.,  {Courteau} S.,    {Gonzalez} J.~J.,  2010,
  submitted to ApJ

\bibitem[\protect\citeauthoryear{{MacArthur}, {Courteau}, {Bell} \&
  {Holtzman}}{{MacArthur} et~al.}{2004}]{Mac04}
{MacArthur} L.~A.,  {Courteau} S.,  {Bell} E.,    {Holtzman} J.~A.,  2004,
  \apjs, 152, 175

\bibitem[\protect\citeauthoryear{{MacArthur}, {Courteau} \&
  {Holtzman}}{{MacArthur} et~al.}{2003}]{Mac03}
{MacArthur} L.~A.,  {Courteau} S.,    {Holtzman} J.~A.,  2003, \apj, 582, 689

\bibitem[\protect\citeauthoryear{{MacArthur}, {Gonz{\'a}lez} \&
  {Courteau}}{{MacArthur} et~al.}{2009}]{Mac09}
{MacArthur} L.~A.,  {Gonz{\'a}lez} J.~J.,    {Courteau} S.,  2009, \mnras, 395,
  28

\bibitem[\protect\citeauthoryear{{Maraston}}{{Maraston}}{2005}]{Mar05}
{Maraston} C.,  2005, \mnras, 362, 799

\bibitem[\protect\citeauthoryear{{Marigo} \& {Girardi}}{{Marigo} \&
  {Girardi}}{2007}]{Mar07}
{Marigo} P.,  {Girardi} L.,  2007, \aap, 469, 239

\bibitem[\protect\citeauthoryear{{Marzke}, {Huchra} \& {Geller}}{{Marzke}
  et~al.}{1994}]{Mar94}
{Marzke} R.~O.,  {Huchra} J.~P.,    {Geller} M.~J.,  1994, \apj, 428, 43

\bibitem[\protect\citeauthoryear{{McDonald}, {Courteau} \& {Tully}}{{McDonald}
  et~al.}{2009a}]{M09c}
{McDonald} M.,  {Courteau} S.,    {Tully} R.~B.,  2009a, \mnras, 393, 628

\bibitem[\protect\citeauthoryear{{McDonald}, {Courteau} \& {Tully}}{{McDonald}
  et~al.}{2009b}]{M09a}
{McDonald} M.,  {Courteau} S.,    {Tully} R.~B.,  2009b, \mnras, 394, 2022

\bibitem[\protect\citeauthoryear{{McDonald}, {Courteau}, {Tully} \&
  {Roediger}}{{McDonald} et~al.}{2010}]{M09b}
{McDonald} M.,  {Courteau} S.,  {Tully} R.~B.,    {Roediger} J.,  2010, \mnras,
  submitted (M10)

\bibitem[\protect\citeauthoryear{{Mehlert}, {Thomas}, {Saglia}, {Bender} \&
  {Wegner}}{{Mehlert} et~al.}{2003}]{Meh03}
{Mehlert} D.,  {Thomas} D.,  {Saglia} R.~P.,  {Bender} R.,    {Wegner} G.,
  2003, \aap, 407, 423

\bibitem[\protect\citeauthoryear{{Mihos} \& {Hernquist}}{{Mihos} \&
  {Hernquist}}{1996}]{MH96}
{Mihos} J.~C.,  {Hernquist} L.,  1996, \apj, 464, 641

\bibitem[\protect\citeauthoryear{{Moore}, {Katz}, {Lake}, {Dressler} \&
  {Oemler}}{{Moore} et~al.}{1996}]{Moo96}
{Moore} B.,  {Katz} N.,  {Lake} G.,  {Dressler} A.,    {Oemler} A.,  1996,
  \nat, 379, 613

\bibitem[\protect\citeauthoryear{{Moorthy} \& {Holtzman}}{{Moorthy} \&
  {Holtzman}}{2006}]{Moo07}
{Moorthy} B.~K.,  {Holtzman} J.~A.,  2006, \mnras, 371, 583

\bibitem[\protect\citeauthoryear{{Navarro}, {Frenk} \& {White}}{{Navarro}
  et~al.}{1995}]{NFW95}
{Navarro} J.~F.,  {Frenk} C.~S.,    {White} S.~D.~M.,  1995, \mnras, 275, 56

\bibitem[\protect\citeauthoryear{{Neistein}, {van den Bosch} \&
  {Dekel}}{{Neistein} et~al.}{2006}]{Nei06}
{Neistein} E.,  {van den Bosch} F.~C.,    {Dekel} A.,  2006, \mnras, 372, 933

\bibitem[\protect\citeauthoryear{{Nelan}, {Smith}, {Hudson}, {Wegner}, {Lucey},
  {Moore}, {Quinney} \& {Suntzeff}}{{Nelan} et~al.}{2005}]{Nel05}
{Nelan} J.~E.,  {Smith} R.~J.,  {Hudson} M.~J.,  {Wegner} G.~A.,  {Lucey}
  J.~R.,  {Moore} S.~A.~W.,  {Quinney} S.~J.,    {Suntzeff} N.~B.,  2005, \apj,
  632, 137

\bibitem[\protect\citeauthoryear{{Nilson}}{{Nilson}}{1973}]{Nil73}
{Nilson} P.,  1973, {Uppsala general catalogue of galaxies}

\bibitem[\protect\citeauthoryear{{Norris}, {Sharples} \& {Kuntschner}}{{Norris}
  et~al.}{2006}]{Nor06}
{Norris} M.~A.,  {Sharples} R.~M.,    {Kuntschner} H.,  2006, \mnras, 367, 815

\bibitem[\protect\citeauthoryear{{Peletier} \& {Balcells}}{{Peletier} \&
  {Balcells}}{1996}]{Pel96}
{Peletier} R.~F.,  {Balcells} M.,  1996, \aj, 111, 2238

\bibitem[\protect\citeauthoryear{{Peletier}, {Balcells}, {Davies},
  {Andredakis}, {Vazdekis}, {Burkert} \& {Prada}}{{Peletier}
  et~al.}{1999}]{Pel99}
{Peletier} R.~F.,  {Balcells} M.,  {Davies} R.~L.,  {Andredakis} Y.,
  {Vazdekis} A.,  {Burkert} A.,    {Prada} F.,  1999, \mnras, 310, 703

\bibitem[\protect\citeauthoryear{{Peletier}, {Falc{\'o}n-Barroso}, {Bacon},
  {Cappellari}, {Davies}, {de Zeeuw}, {Emsellem} \& {plus 6 more}}{{Peletier}
  et~al.}{2007}]{Pel07}
{Peletier} R.~F.,  {Falc{\'o}n-Barroso} J.,  {Bacon} R.,  {Cappellari} M.,
  {Davies} R.~L.,  {de Zeeuw} P.~T.,  {Emsellem} E.,    {plus 6 more} 2007,
  \mnras, 379, 445

\bibitem[\protect\citeauthoryear{{Poggianti}, {Bridges}, {Carter}, {Mobasher},
  {Doi}, {Iye}, {Kashikawa} \& {plus 6 more}}{{Poggianti} et~al.}{2001}]{Pog01}
{Poggianti} B.~M.,  {Bridges} T.~J.,  {Carter} D.,  {Mobasher} B.,  {Doi} M.,
  {Iye} M.,  {Kashikawa} N.,    {plus 6 more} 2001, \apj, 563, 118

\bibitem[\protect\citeauthoryear{{Portinari}, {Sommer-Larsen} \&
  {Tantalo}}{{Portinari} et~al.}{2004}]{Por04}
{Portinari} L.,  {Sommer-Larsen} J.,    {Tantalo} R.,  2004, \mnras, 347, 691

\bibitem[\protect\citeauthoryear{{Quilis}, {Moore} \& {Bower}}{{Quilis}
  et~al.}{2000}]{Qui00}
{Quilis} V.,  {Moore} B.,    {Bower} R.,  2000, Science, 288, 1617

\bibitem[\protect\citeauthoryear{{Rawle}, {Smith} \& {Lucey}}{{Rawle}
  et~al.}{2010}]{Raw09}
{Rawle} T.~D.,  {Smith} R.~J.,    {Lucey} J.~R.,  2010, \mnras, 401, 852

\bibitem[\protect\citeauthoryear{{Richard}, {Brook}, {Martel}, {Kawata},
  {Gibson} \& {Sanchez-Blazquez}}{{Richard} et~al.}{2010}]{Rich09}
{Richard} S.,  {Brook} C.~B.,  {Martel} H.,  {Kawata} D.,  {Gibson} B.~K.,
  {Sanchez-Blazquez} P.,  2010, \mnras, 402, 1489

\bibitem[\protect\citeauthoryear{{Rickes}, {Pastoriza} \& {Bonatto}}{{Rickes}
  et~al.}{2009}]{Ric09}
{Rickes} M.~G.,  {Pastoriza} M.~G.,    {Bonatto} C.,  2009, ArXiv e-prints

\bibitem[\protect\citeauthoryear{{Roediger}, {Courteau}, {McDonald} \&
  {MacArthur}}{{Roediger} et~al.}{2010}]{Roe10}
{Roediger} J.,  {Courteau} S.,  {McDonald} M.,    {MacArthur} L.,  2010,
  submitted for publication

\bibitem[\protect\citeauthoryear{{Ro{\v s}kar}, {Debattista}, {Quinn},
  {Stinson} \& {Wadsley}}{{Ro{\v s}kar} et~al.}{2008}]{Ros08b}
{Ro{\v s}kar} R.,  {Debattista} V.~P.,  {Quinn} T.~R.,  {Stinson} G.~S.,
  {Wadsley} J.,  2008, \apjl, 684, L79

\bibitem[\protect\citeauthoryear{{Ro{\v s}kar}, {Debattista}, {Stinson},
  {Quinn}, {Kaufmann} \& {Wadsley}}{{Ro{\v s}kar} et~al.}{2008}]{Ros08}
{Ro{\v s}kar} R.,  {Debattista} V.~P.,  {Stinson} G.~S.,  {Quinn} T.~R.,
  {Kaufmann} T.,    {Wadsley} J.,  2008, \apjl, 675, L65

\bibitem[\protect\citeauthoryear{{S{\'a}nchez-Bl{\'a}zquez}, {Forbes},
  {Strader}, {Brodie} \& {Proctor}}{{S{\'a}nchez-Bl{\'a}zquez}
  et~al.}{2007}]{San07}
{S{\'a}nchez-Bl{\'a}zquez} P.,  {Forbes} D.~A.,  {Strader} J.,  {Brodie} J.,
  {Proctor} R.,  2007, \mnras, 377, 759

\bibitem[\protect\citeauthoryear{{S{\'a}nchez-Bl{\'a}zquez}, {Gorgas},
  {Cardiel} \& {Gonz{\'a}lez}}{{S{\'a}nchez-Bl{\'a}zquez} et~al.}{2006}]{San06}
{S{\'a}nchez-Bl{\'a}zquez} P.,  {Gorgas} J.,  {Cardiel} N.,    {Gonz{\'a}lez}
  J.~J.,  2006, \aap, 457, 809

\bibitem[\protect\citeauthoryear{{Schechter}}{{Schechter}}{1976}]{Sch76}
{Schechter} P.,  1976, \apj, 203, 297

\bibitem[\protect\citeauthoryear{{Schiavon}}{{Schiavon}}{2007}]{Sch07}
{Schiavon} R.~P.,  2007, \apjs, 171, 146

\bibitem[\protect\citeauthoryear{{Schiavon}, {Faber}, {Rose} \&
  {Castilho}}{{Schiavon} et~al.}{2002}]{Sch02}
{Schiavon} R.~P.,  {Faber} S.~M.,  {Rose} J.~A.,    {Castilho} B.~V.,  2002,
  \apj, 580, 873

\bibitem[\protect\citeauthoryear{{Schlegel}, {Finkbeiner} \&
  {Davis}}{{Schlegel} et~al.}{1998}]{Sch98}
{Schlegel} D.~J.,  {Finkbeiner} D.~P.,    {Davis} M.,  1998, \apj, 500, 525

\bibitem[\protect\citeauthoryear{{Serra}, {Trager}, {Oosterloo} \&
  {Morganti}}{{Serra} et~al.}{2008}]{Ser08}
{Serra} P.,  {Trager} S.~C.,  {Oosterloo} T.~A.,    {Morganti} R.,  2008, \aap,
  483, 57

\bibitem[\protect\citeauthoryear{{Shapley}, {Steidel}, {Erb}, {Reddy},
  {Adelberger}, {Pettini}, {Barmby} \& {Huang}}{{Shapley} et~al.}{2005}]{Sha05}
{Shapley} A.~E.,  {Steidel} C.~C.,  {Erb} D.~K.,  {Reddy} N.~A.,  {Adelberger}
  K.~L.,  {Pettini} M.,  {Barmby} P.,    {Huang} J.,  2005, \apj, 626, 698

\bibitem[\protect\citeauthoryear{{Sil'chenko}}{{Sil'chenko}}{2006}]{Sil06}
{Sil'chenko} O.~K.,  2006, \apj, 641, 229

\bibitem[\protect\citeauthoryear{{Skrutskie}, {Cutri}, {Stiening}, {Weinberg},
  {Schneider}, {Carpenter}, {Beichman} \& {plus 24 more}}{{Skrutskie}
  et~al.}{2006}]{Skr06}
{Skrutskie} M.~F.,  {Cutri} R.~M.,  {Stiening} R.,  {Weinberg} M.~D.,
  {Schneider} S.,  {Carpenter} J.~M.,  {Beichman} C.,    {plus 24 more} 2006,
  \aj, 131, 1163

\bibitem[\protect\citeauthoryear{{Smith}, {Lucey} \& {Hudson}}{{Smith}
  et~al.}{2009}]{Smi09}
{Smith} R.~J.,  {Lucey} J.~R.,    {Hudson} M.~J.,  2009, \mnras, 400, 1690

\bibitem[\protect\citeauthoryear{{Steinmetz} \& {Navarro}}{{Steinmetz} \&
  {Navarro}}{2002}]{Ste02}
{Steinmetz} M.,  {Navarro} J.~F.,  2002, New Astronomy, 7, 155

\bibitem[\protect\citeauthoryear{{Strateva}, {Ivezi{\'c}}, {Knapp},
  {Narayanan}, {Strauss}, {Gunn}, {Lupton} \& {plus 22 more}}{{Strateva}
  et~al.}{2001}]{Str01}
{Strateva} I.,  {Ivezi{\'c}} {\v Z}.,  {Knapp} G.~R.,  {Narayanan} V.~K.,
  {Strauss} M.~A.,  {Gunn} J.~E.,  {Lupton} R.~H.,    {plus 22 more} 2001, \aj,
  122, 1861

\bibitem[\protect\citeauthoryear{{Tamura} \& {Ohta}}{{Tamura} \&
  {Ohta}}{2003}]{Tam03}
{Tamura} N.,  {Ohta} K.,  2003, \aj, 126, 596

\bibitem[\protect\citeauthoryear{{Taylor}, {Jansen}, {Windhorst}, {Odewahn} \&
  {Hibbard}}{{Taylor} et~al.}{2005}]{Tay05}
{Taylor} V.~A.,  {Jansen} R.~A.,  {Windhorst} R.~A.,  {Odewahn} S.~C.,
  {Hibbard} J.~E.,  2005, \apj, 630, 784

\bibitem[\protect\citeauthoryear{{Thomas}, {Maraston}, {Bender} \& {Mendes de
  Oliveira}}{{Thomas} et~al.}{2005}]{Tho05}
{Thomas} D.,  {Maraston} C.,  {Bender} R.,    {Mendes de Oliveira} C.,  2005,
  \apj, 621, 673

\bibitem[\protect\citeauthoryear{{Tikhonov}, {Galazutdinova} \&
  {Aparicio}}{{Tikhonov} et~al.}{2003}]{Tik03}
{Tikhonov} N.~A.,  {Galazutdinova} O.~A.,    {Aparicio} A.,  2003, \aap, 401,
  863

\bibitem[\protect\citeauthoryear{{Tonini}, {Maraston}, {Devriendt}, {Thomas} \&
  {Silk}}{{Tonini} et~al.}{2009}]{Ton09}
{Tonini} C.,  {Maraston} C.,  {Devriendt} J.,  {Thomas} D.,    {Silk} J.,
  2009, \mnras, 396, L36

\bibitem[\protect\citeauthoryear{{Trager}, {Faber}, {Worthey} \&
  {Gonz{\'a}lez}}{{Trager} et~al.}{2000}]{Tra00}
{Trager} S.~C.,  {Faber} S.~M.,  {Worthey} G.,    {Gonz{\'a}lez} J.~J.,  2000,
  \aj, 119, 1645

\bibitem[\protect\citeauthoryear{{Tremonti}, {Heckman}, {Kauffmann},
  {Brinchmann}, {Charlot}, {White}, {Seibert} \& {plus 5 more}}{{Tremonti}
  et~al.}{2004}]{Tre04}
{Tremonti} C.~A.,  {Heckman} T.~M.,  {Kauffmann} G.,  {Brinchmann} J.,
  {Charlot} S.,  {White} S.~D.~M.,  {Seibert} M.,    {plus 5 more} 2004, \apj,
  613, 898

\bibitem[\protect\citeauthoryear{{van den Bergh}}{{van den
  Bergh}}{1994}]{Van94}
{van den Bergh} S.,  1994, \aj, 107, 153

\bibitem[\protect\citeauthoryear{{Vazdekis}}{{Vazdekis}}{1999}]{Vaz99}
{Vazdekis} A.,  1999, \apj, 513, 224

\bibitem[\protect\citeauthoryear{{Vazdekis}, {Casuso}, {Peletier} \&
  {Beckman}}{{Vazdekis} et~al.}{1996}]{Vaz96}
{Vazdekis} A.,  {Casuso} E.,  {Peletier} R.~F.,    {Beckman} J.~E.,  1996,
  \apjs, 106, 307

\bibitem[\protect\citeauthoryear{{Vazdekis}, {Kuntschner}, {Davies}, {Arimoto},
  {Nakamura} \& {Peletier}}{{Vazdekis} et~al.}{2001}]{Vaz01}
{Vazdekis} A.,  {Kuntschner} H.,  {Davies} R.~L.,  {Arimoto} N.,  {Nakamura}
  O.,    {Peletier} R.,  2001, \apjl, 551, L127

\bibitem[\protect\citeauthoryear{{Weinzirl}, {Jogee}, {Khochfar}, {Burkert} \&
  {Kormendy}}{{Weinzirl} et~al.}{2009}]{Wei09}
{Weinzirl} T.,  {Jogee} S.,  {Khochfar} S.,  {Burkert} A.,    {Kormendy} J.,
  2009, \apj, 696, 411

\bibitem[\protect\citeauthoryear{{Wiklind} \& {Henkel}}{{Wiklind} \&
  {Henkel}}{2001}]{Wik01}
{Wiklind} T.,  {Henkel} C.,  2001, \aap, 375, 797

\bibitem[\protect\citeauthoryear{{Worthey}}{{Worthey}}{1994}]{Wor94}
{Worthey} G.,  1994, \apjs, 95, 107

\bibitem[\protect\citeauthoryear{{Yoachim} \& {Dalcanton}}{{Yoachim} \&
  {Dalcanton}}{2006}]{Yoa08}
{Yoachim} P.,  {Dalcanton} J.~J.,  2006, \aj, 131, 226

\bibitem[\protect\citeauthoryear{{York}, {Adelman}, {Anderson} Jr., {Anderson},
  {Annis}, {Bahcall}, {Bakken} \& {plus 137 more}}{{York} et~al.}{2000}]{Yor00}
{York} D.~G.,  {Adelman} J.,  {Anderson} Jr. J.~E.,  {Anderson} S.~F.,  {Annis}
  J.,  {Bahcall} N.~A.,  {Bakken} J.~A.,    {plus 137 more} 2000, \aj, 120,
  1579

\bibitem[\protect\citeauthoryear{{Zackrisson}, {Bergvall}, {{\"O}stlin},
  {Micheva} \& {Leksell}}{{Zackrisson} et~al.}{2006}]{Zak06}
{Zackrisson} E.,  {Bergvall} N.,  {{\"O}stlin} G.,  {Micheva} G.,    {Leksell}
  M.,  2006, \apj, 650, 812

\bibitem[\protect\citeauthoryear{{Zibetti}, {White} \& {Brinkmann}}{{Zibetti}
  et~al.}{2004}]{Zib04}
{Zibetti} S.,  {White} S.~D.~M.,    {Brinkmann} J.,  2004, \mnras, 347, 556

\end{thebibliography}
